\documentclass[fleqn,usenatbib]{mnras}

\usepackage{newtxtext,newtxmath}

\usepackage[T1]{fontenc}
\usepackage{ae,aecompl}

\usepackage{graphicx}	
\usepackage{subfig}
\usepackage{amsmath}	
\usepackage{float}
\usepackage{comment}
\usepackage{booktabs}
\usepackage{wrapfig}
\usepackage{xcolor}

\usepackage{longtable}
\newcommand*{\courier}{\fontfamily{<pcr>}\selectfont}

\title[Asteroidal occultations]{Precise astrometry and diameters of asteroids from occultations -- a data-set of observations and their interpretation}

\author[David Herald et al.]{David Herald $^{1}$\thanks{E-mail:drherald@bigpond.net.au (DRH)}, 
David Gault$^{2}$,
Robert Anderson$^{3}$,
David Dunham$^{4}$,
Eric Frappa$^{5}$,
\newauthor Tsutomu Hayamizu$^{6}$, Steve Kerr$^{7}$, Kazuhisa Miyashita$^{8}$, John Moore$^{9}$,
\newauthor Hristo Pavlov$^{10}$, Steve Preston$^{11}$, John Talbot$^{12}$, Brad Timerson (deceased)$^{13}$
\\
$^{1}$Trans Tasman Occultation Alliance, drherald@bigpond.net.au\\
$^{2}$Trans Tasman Occultation Alliance, davegault@bigpond.com\\
$^{3}$International Occultation Timing Association, bob.anderson.ok@gmail.com\\
$^{4}$International Occultation Timing Association, dunham@starpower.net\\
$^{5}$Euraster, frappa@laposte.net\\
$^{6}$Japanese Occultation Information Network, haya3zustar@gmail.com\\
$^{7}$Trans Tasman Occultation Alliance, Steve.Kerr@outlook.com.au\\
$^{8}$Japanese Occultation Information Network, k\_miyash@nifty.com\\
$^{9}$International Occultation Timing Association, john@jmooreou.com\\
$^{10}$International Occultation Timing Association -- European Section,  hristo\_dpavlov@yahoo.com\\
$^{11}$International Occultation Timing Association, stevepr@netstevepr.com\\
$^{12}$Trans Tasman Occultation Alliance, john.talbot@xtra.co.nz\\
$^{13}$International Occultation Timing Association, deceased\\
}

\date{Accepted XXX. Received YYY; in original form ZZZ}

\pubyear{2020}

\begin{document}
\label{firstpage}
\pagerange{\pageref{firstpage}--\pageref{lastpage}}
\maketitle

\begin{abstract}
Occultations of stars by asteroids have been observed since 1961, increasing from a very small number to now over 500 annually. We have created and regularly maintain a growing data-set of more than 5,000 observed asteroidal occultations. The data-set includes: the raw observations; astrometry at the 1 mas level based on centre of mass or figure (not illumination); where possible the asteroid's diameter to 5 km or better, and fits to shape models; the separation and diameters of asteroidal satellites; and double star discoveries with typical separations being in the tens of mas or less. The data-set is published at NASA's Planetary Data System and is regularly updated. We provide here an overview of the data-set, discuss the issues associated with determining the astrometry and diameters, and give examples of what can be derived from the data-set. We also compare the occultation diameters of asteroids with the diameters measured by the satellites \textit{NEOWISE}, \textit{AKARI AcuA}, and \textit{IRAS}, and show that the best satellite-determined diameter is a combination of the diameters from all three satellites.
\end{abstract}




\section{Introduction}

Occultations by asteroids (asteroidal occultations) provide one of the few means of directly measuring the size and shape of an asteroid. It can be applied to any asteroid, and does not require access to a large telescope to image the asteroid. All that is required is an imaging system that can detect the star that will be occulted, the observer's coordinates, and accurate measurement of the time of disappearance and reappearance of the star. Desirably several observers are located across the occultation path to determine the shape and size of the asteroid.

Most asteroidal occultations have been observed by amateur astronomers using small-aperture telescopes. Additionally regular campaigns on objects of high astrophysical interest are initiated by professional programs, including the ERC Lucky Star\footnote{\textcolor{blue}{https://lesia.obspm.fr/lucky-star/}} project of l'Observatoire de Paris and the RECON\footnote{\textcolor{blue}{http://tnorecon.net}} project.  Early efforts were largely visual observations using techniques then common with lunar occultation observations. However evidence arose that the Personal Equation correction  (the visual observer's reaction time to an event) required for asteroidal occultations was generally much greater than for lunar occultations \citep{1990AJ.....99.1636D}. 

With early efforts, the large prediction uncertainties required the observer to visually monitor the star for many minutes. This undoubtedly led to a number of false events -- either from tiredness or atmospherics. The data-set of asteroidal occultation observations discussed in this paper has been thoroughly reviewed to identify such events.

In the late 1990's, observers started experimenting with inexpensive Composite Video Base-band Signal (CVBS) video cameras (PAL and NTSC). Devices were subsequently developed to insert time into the video stream, using the GPS 1 pulse-per-second signal as the time reference. Software to measure such videos and the associated time stamps was also developed. These systems ensure event times are accurate to the frame-rate of the video recording being used. They also avoid the issue of false events from tiredness. IOTA's Observing Manuals\footnote{\textcolor{blue}{http://occultations.org/observing/educational-materials/observing-manuals/}} provides an overview of current techniques.

Many observations are made from an observer's home location. Some observers have mobile capability, and travel to the location of the predicted path. Observing locations are coordinated using the free software {\scshape OccultWatcher}{\footnote{\textcolor{blue}{https://www.occultwatcher.net}}}. Some observers have built multiple portable recording systems; before an event, they are deployed along roadsides behind vegetation, prepointed to where the star will be at the time of the occultation; after the event they are collected as the observer returns home. Such arrays have provided some of the best occultation profiles.

As a result of these developments, the reliability of reported observations has increased from 50 per cent in 1990 to better than 96 per cent from 2008. At the same time the annual number of reported events has increased from about 10 in 1990 to 586 in 2019.

The biggest challenge remains the ephemeris uncertainty for the asteroid. For a main-belt asteroid, an angular displacement of 1~mas is crudely equivalent to a displacement of the asteroid's shadow of about 1~km. The ephemeris uncertainty for most main belt asteroids (as at 2019) is generally greater than about 40~mas, resulting in uncertainty in the path location generally being greater than 40~km. As \textit{Gaia} asteroid astrometry is incorporated into orbit solutions, we expect the uncertainty in the path location of main belt asteroids will reduce to the kilometre level.

\section{The data-set of events and observations}
The data-set of asteroidal occultation observations is published at NASA's Planetary Data System, Small bodies node, Asteroid//Dust Subnode\footnote{\textcolor{blue}{https://sbn.psi.edu/pds/resource/occ.html\label{PDS}}}, from which it can be browsed or downloaded. It includes main belt asteroids, Trojans, Centaurs, TNO's and NEO's.  The data-set is updated on an approximate annual basis.

There are 3 groups of files that makeup the archive:
\begin{itemize} 
    \item Events involving asteroids. \textbf{\textit{occlist\_2019.tab}} contains the overall data for each event, while \textbf{\textit{occtimings\_2019.tab}} contains the data for each observer.
    \item Events involving a major planet, or the satellite of a major planet. \hspace{1mm} \textbf{\textit{occsatlist\_2019.tab}} contains the overall data for each event, while \textbf{\textit{occsattimings\_2019.tab}} contains the data for each observers.
    \item A summary of the more important data for each event is held in the files \textbf{\textit{occsummary\_2019.tab}} and \textbf{\textit{occsatsummary\_2019.tab}}
\end{itemize}

Changes in observing methods, timing equipment, star catalogues, and information about asteroids, has resulted in an evolving data-set. For example, the current data-set (V3.0) contains for the first time a matching of occultation results to asteroid shape models. As the data-set evolves, it provides an ongoing basis for researchers to access the observations and analyse them in a manner that best suits their needs.

The \textbf{\textit{bundle\_description.txt}} file that is part of the archive provides information about how stars are identified, shape models, and the content of the various files in the archive. Importantly, it contains a comprehensive list of issues that may affect the data in the data-set. Users should make sure they are familiar with this file.

[Additionally, about 8000 light curves from asteroidal and lunar occultations are available at VizieR, catalogue B/occ.\footnote{\color{blue}{https://vizier.u-strasbg.fr/viz-bin/VizieR?-source=B/occ}\label{VizieR}} ]

Amateur observers of the occultation community generally use the software package called {\scshape Occult}\footnote{\textcolor{blue}{http://www.lunar-occultations.com/iota/occult4.html}}. Written for the Windows{\textsuperscript{®}} environment, it provides extensive tools for analyzing asteroidal occultations, as well as computing predictions. The data-set at NASA's Planetary Data System \textsuperscript{\ref{PDS}} is created on an approximately annual basis from the observations file maintained by the {\scshape Occult} software. Updates to that observations file occur on an approximately monthly basis, with users being alerted to the presence of updates from within the software (which also handles the download). For those not using {\scshape Occult}, the observations file can be separately downloaded at any time\footnote{\textcolor{blue}{http://www.lunar-occultations.com/occult4/asteroid\_observations.zip}}. The downloaded file holds the observations in an XML-like file, together with an .htm file that provides the file format. 

The data-set of asteroidal occultations contains over 15,000 observations by more than 3,300 individuals from around the world over a period of more than 40 years. It involves more than 1,330 individual asteroids observed in 4,400 separate occultations. The great majority of observers have made these observations at their own expense. Users of this data-set are requested to acknowledge their contributions with a statement like:\newline
\phantom{l}
\hrule
\phantom{l}\newline
{\addtolength{\leftskip}{5mm}We acknowledge the contributions of the over 3,000 observers who have provided the observations in the data-set. Most of those observers are affiliated with one or more of:
\begin{itemize}
    \item European Asteroidal Occultation Network (EAON)
    \item International Occultation Timing Association (IOTA)
    \item International Occultation Timing Association -- European Section (IOTA--ES)
    \item Japanese Occultation Information Network (JOIN)
    \item Trans Tasman Occultation Alliance (TTOA)
\end{itemize}
}
\hrule

\section{Quality assurance}
The collection of observation reports from a large demographic requires a rigorous quality assurance regime.  Prior to the year 2000 there was little or no quality assurance, leading to a large number of unreliable observations being incorporated into the data-set, however the entire data-set has been reviewed to identify, flag, and exclude such observations from any further analysis.  The quality assurance regime that has evolved since the year 2000 is described as follows:

\begin{itemize}
    \item Most occultation observation timings are extracted from the recording using one of three well established tools:
    \begin{itemize}
        \item {\scshape Limovie}\footnote{\textcolor{blue}{http://astro-limovie.info/limovie/limovie\_en.html}}, 
       \item {\scshape PyMovie}\footnote{\textcolor{blue}{http://occultations.org/observing/software/pymovie/}}
       \item {\scshape Tangra}\footnote{\textcolor{blue}{http://www.hristopavlov.net/Tangra3}}
   \end{itemize}
    
    Each produces a light curve that graphically demonstrates the characteristics of the observation, and can be analysed numerically.  
    
    \item Statistical analysis of the observations is commonly aided by the use of several tools:
      \begin{itemize} 
          \item {\scshape AOTA}, in {\scshape Occult} \footnote{\textcolor{blue}{http://www.lunar-occultations.com/iota/occult4.html}}
          \item {\scshape Occular}, {\scshape ROTE} and {\scshape PYOTE}\footnote{\textcolor{blue}{http://occultations.org/observing/software/ote/}} 
    \end{itemize}
    
    \item Structured report forms are used to ensure all necessary information is collected.  Many analysis tools populate the report forms automatically, thereby reducing the chances of transcription errors.
    
    \item The world is divided into a number of regions -- Australasia, Europe, Japan and North America, together with South America and India. Observers in those regions submit their reports to a regional coordinator for collation. The regional coordinator examines each report for typographical errors, completeness, inconsistencies with other observers, or anything that might be questionable. In appropriate circumstances the coordinator will ask for independent review or analysis of the observation. A situation where this might occur is exemplified in the discussion at the end of section \ref{486958}, of an occultation by (486958) Arrokoth.
    
    \item The coordinator will also add a ‘prediction’ point to the set of observations. This is particularly important for single-chord observations, which cannot be confirmed by consistency with another observed chord. A major discrepancy between an observed chord and the prediction point may lead to the observation being given no weight, or disregarded.
    
    \item Once complete, the set of observations are forwarded to a global coordinator, who adds a fit to shape models (when available) and looks for any anomalies in observations.  (In recent years the main anomalies have been associated with the use of NTP as a time source, personal equation for visual observations, and observers attempting an event at the limit of their system's capabilities.) The set of observations is then entered into the data-set maintained by the software {\scshape Occult} \textit{supra}, from which the data-set maintained at NASA's Planetary Data System\textsuperscript{\ref{PDS}} is periodically updated.
\end{itemize}
The entire data-set of observations was reviewed prior to its archiving at NASA's Planetary Data System\textsuperscript{\ref{PDS}} as Version 3.0. In that review, ‘post-dictions’ were generated for all events prior to 2000 to identify unreliable observations. Post-dictions are in the process of being added for all single chord events after 2000 up until when their inclusion became routine.

\section{Reduction of the observations}\label{reductions}
The fundamental observation made in an asteroidal occultation is the time (UTC) of the star disappearing and reappearing as the asteroid passes in front of it -- as observed from a site on the Earth identified by geographic coordinates. Ideally observations are made at several sites. Events may not be a simple disappear and reappear. There may be:
\begin{itemize}
   \item step events caused by the star being double;
   \item multiple events caused by
   \begin{itemize}
       \item the star being a relatively wide double;
       \item the asteroid having one or more satellites or one or more rings; 
       \item the star passing behind two or more topographic features on the asteroid's limb. 
   \end{itemize}
   \item gradual events caused by the star having an appreciable apparent diameter, or the effects of Fresnel diffraction.
\end{itemize}

The goal of the reduction process is to combine all observations such that the position and shape of the asteroid is well defined. This requires referring all observations to a common time on a plane that is normal to the asteroid's shadow (the fundamental plane). This involves:
\begin{itemize}
    \item using Besselian elements;
    \item using the \textit{apparent} positions of both the star and asteroid to define the orientation of the fundamental plane, to ensure the geographic position of each observer at their event times is correctly located on the fundamental plane;
    \item computing a cubic expression representing the motion of the asteroid's shadow on the fundamental plane; and
    \item using the cubic expression to move the observer locations on the fundamental plane for each event time to their location at a common reference time.
\end{itemize}

The reference time is determined on the following basis:
\begin{itemize}
    \item[\textbf{a.}] for occultations that do not involve satellites or a double star, the mean time of all disappearance and reappearance events used in the analysis.
    \item[\textbf{b.}] if a double star is involved, the mean time of all events involving the primary component.
    \item[\textbf{c.}] for occultations involving a satellite:
    \begin{itemize}
         \item[\textbf{c1.}] For events involving a relatively small satellite, the largely different masses result in the displacement from orbital motion being much greater for the satellite than for the main body. The reference time is set as the mean time of all disappearance and reappearance events used in the analysis of the satellite. Events involving the main body are moved to this reference time, with the motion of the main body treated as being that of the system as a whole. This eliminates the effects of orbital motion from the measurement of relative positions.
         
        \item[\textbf{c2.}]For binary asteroids having components of similar size and relative proximity [such as (90) Antiope], the mean time of all events used in the analysis. Any time differences between the two components will usually be small, such that issues associated with orbital motion are insignificant. The main difficulty is identifying the location of the centre of mass of such a system, for reporting precise astrometry.
    \end{itemize}
\end{itemize}
After the position of each event has been moved to a common reference time, those positions map out the location and shape of the asteroid's shadow on the fundamental plane and (where relevant) the relative positions of a parent body and its satellite, and the displacement between the separate shadows arising from a double star. A shape model or ellipse fitted to those positions enables the size of the asteroid, and the position of its centre, to be determined.

The fitting of an ellipse to the observed chords involves the free parameters of the x,y coordinates of the centre of the ellipse, the major and minor axes of the fitted ellipse, and its position angle. If a double star is involved, the separation and position angle of the companion star are also included as free parameters. The residuals are computed along the radius of the ellipse. Miss observations are only considered in the iteration when the miss chord intersects the currently-evaluated ellipse solution. When that occurs, the chord is included in the solution as an event point, but with its distance from the centre of the ellipse being reduced by 0.5 km. The resulting fit will either miss this chord, or intersect it by only a small distance (consistent with the overall uncertainties in the observations and asteroid shape). When fitting a circle to the observations, the free parameters for the minor axis and orientation of the ellipse are excluded from the solution.

For events having insufficient chords to derive the asteroid's size, the location of the centre of the asteroid relies on fitting a circle having an assumed diameter to the chords. The assumed diameter was derived whenever possible as a weighted mean of all the individual measures from each of the following satellites:
\begin {itemize}
\item \textit{IRAS} \citep{Tedesco2004}
\item \textit{AKARI AcuA} \citep{AliLagoa2018} 
\item \textit{NEOWISE} Mainbelt, Centaurs, Hildas, Jupiter\_Trojans, Neos \citep{Neowise}
\item The midcourse space experiment thermal infrared minor planet survey \citep{Tedesco2002}
\end{itemize}
Where an asteroid was not included in any of these satellites, the diameter was estimated from the asteroid's absolute magnitude H using \begin{math} Diameter = 10^{(3.1295 - 0.5\log_{10}(a) - 0.2H)}\end{math} with an assumed albedo of a=0.1652 \textendash{} \citep{1989aste.conf..524B} at p. 551.

The assumed diameter so calculated, and its uncertainty, is included in the data-set for each event -- irrespective of whether the chords are sufficient to derive the asteroid's size.

\section{Astrometry}\label{astrometry}
Astrometry from an asteroidal occultation is derived from the offset of the asteroid from the star at the reference time. For all but 10 events the position of the star is from \textit{Gaia} DR2. Therefore all but 10 asteroid positions are on the \textit{Gaia} DR2 reference frame. For eight events, the position of the star is from Hipparcos2\footnote{\textcolor{blue}{https://cdsarc.unistra.fr/viz-bin/cat/I/311}} \citep{Hipparcos}. For the remaining two events, the position of the star is from UCAC4\footnote{\textcolor{blue}{https://cdsarc.unistra.fr/viz-bin/cat/I/322A}} \citep{ucac4}.

The method used to determine the position of the asteroid is:
\begin{itemize}
\item determine the (x,y) coordinates of the centre (however determined) of the shadow on the fundamental plane at the reference time;

\item conduct an ‘inverse’ Besselian Element calculation to derive the offset from the star's apparent position;

\item convert the offset in the apparent reference frame to an offset in the J2000 ICRS reference frame -- by removing the effects of precession, nutation, and aberration;

\item add the offset to the J2000 ICRS position of the star as corrected for parallax, and for proper motion (including foreshortening -- which arises when the distance to a star having a large parallax is significantly different from its distance at the catalogue epoch; it is negligible for the majority of stars in our data-set).

\item The \textit{astrometric} position of the asteroid is obtained by correcting for relativistic deflection by the Sun of the asteroid relative to the star. This can be significant for solar elongations as large as 145$^{\circ}$ or more. 
\end{itemize}

The data-set contains the J2000 ICRS position of the star corrected for proper motion (including foreshortening) and parallax, and the offset of the asteroid from that star position -- the combination of which gives the J2000 ICRS position of the asteroid. Also provided is the deflection of the asteroid relative to the star, needed to obtain its astrometric position.

The data-set contains astrometry for over 4,000 asteroids. The accuracy of those positions varies from less than one mas to tens of mas, depending \textit{inter alia} on whether a shape model is available, multiple chords are observed, or (for single chord observations) the angular diameter of the asteroid. Astrometry for 19 TNO's and 4 Centaurs are derived by \textcolor{blue}{Rommel et al (submitted to A\&A)}; for those included in our data-set, the positions are essentially identical. 

\subsection{Uncertainties in the astrometry} \label{Uncertainties}
 Under the right circumstances a well-observed occultation of a main-belt asteroid can establish the position of the shadow to better than 1~km -- equivalent to less than 0.5~mas for a main-belt asteroid. Factors relevant to the actual precision achievable are:

\begin{itemize}
    \item Events where only one observer records an occultation, or there are several observers but their chords are closely spaced. This is a common situation. The astrometric position is based on positioning the asteroid on the centre of the observed chord(s). Since the asteroid could be on one or other side of the chord, the across-path uncertainty is set at 80 per cent of the apparent assumed radius of the asteroid. The along-path uncertainty is better determined. If the chord length is similar to the expected diameter of the asteroid, the mid-point of that chord will approximately correspond to the along-path centre of the asteroid. However if the chord length is considerably less than the expected duration, the possibility of the asteroid's profile being significantly non-spherical must be considered -- potentially displacing the along-path centre of the asteroid from the mid-point of the chord. Accordingly the along-path uncertainty is estimated as: 
    \begin{itemize} 
      \item[\textbf{a.}] chord length > 80 per cent of the asteroid's assumed diameter -- 5 per cent of that diameter;
      \item[\textbf{b.}] chord length between 60 per cent and 80 per cent of the asteroid's assumed diameter -- 10 per cent of that diameter;
      \item[\textbf{c.}] chord length < 60 per cent of the asteroid's assumed diameter -- 20 per cent of that diameter;
    \end{itemize}
    This issue is not dealt with in V.3.0 of the data-set, but will be in subsequent versions.
    
    \item Events where there are only a small number of chords, all located on one half of the asteroid. These events are fitted using a least-squares best fit to a circle having the assumed diameter of the asteroid. However if the distribution of the chords permit, the diameter of the asteroid is included as a free parameter. In some situations the observed chords are displaced in a manner that requires a fit to an ellipse. That ellipse is usually subjected to the arbitrary constraint of its area being that of a circle having the assumed diameter of the asteroid.
    
    \item Events where there is a good distribution of chords. For these a least-squares solution where the location of the centre of the ellipse, the dimensions of the major and minor axes of the ellipse, and its orientation as free parameters. If the event involves a double star, the position angle and separation of the companion are also free parameters.
    
    \item Events where the occultation chords provide a good fit to a shape model. Astrometry is derived using the centre of that shape model, which is coincident with the centre of mass of the asteroid (assuming uniform density). The uncertainty in the astrometry is set (somewhat arbitrarily) at 0.02 of the asteroid's diameter -- reflecting the uncertainties in both the shape model and the occultation chords, yet recognizing the close matching between the chords and shape model.
\end{itemize}

Where a least squares solution is undertaken, there are occasions where one or several variables cannot be treated as free parameters -- to prevent divergence when iterating for the best-fit solution. The parameters that are incorporated in the least squares solution are listed in the data-set.

For the least squares solution, individual observations are weighted. Some observations have specific weights applied. However almost all observations use a default weighting based on the method of observation. Currently the default weighting is 5 for analogue video, 4 for digital video or photometer, 3 for sequential images (e.g. with a CCD camera) or a drift-scan (where the star field drifts across a CCD imager), 1 for visual and any other types of observation. A weighting of 0 indicates the event has been excluded from the solution. These weightings will likely change as observing techniques evolve over time.

\subsection{Uncertainties not included in the current data-set} 
The data-set is in a state of continuous evolution. The uncertainties provided in the current data-set are limited to the fitting of the ellipse (or circle) to the observed chords. Other sources of uncertainty that will be included in future releases of the data-set are:
\begin{itemize}
\item Uncertainty in the star positions. For stars from \textit{Gaia} DR2 and Hipparcos2, the uncertainties are in right ascension, declination, proper motion, and parallax. For the two stars with positions from UCAC4 \citep{ucac4}, there is no parallax.
\item For \textit{Gaia} DR2, stars brighter than about magnitude 12.5 are affected by a frame rotation error \citep{2018A&A...616A...2L} at section 5.1, and \citep{2020A&A...633A...1L}.
\item The event times reported by each observer have an associated uncertainty. In V3.0 these uncertainties, which primarily affect the reference time, are not considered -- but will be incorporated in future versions. However many past observations have no reported time uncertainties. For these observations uncertainties will be set (depending on the mode of observation and time source used) at one of 0.5, 1.0 or 1.5 seconds, with weights of 3, 2 or 1.
\end{itemize}

\subsection{Particular characteristics}
Astrometry from an occultation event differs from traditional astrometry:
\begin{itemize}
\item The astrometry is associated with a single star. Consequently it can be updated whenever the star position is updated. This is impractical for traditional astrometry reliant on plate solutions involving many stars.

\item The orbital motion of an asteroid is dictated by its centre of mass, not illumination nor figure. Traditional astrometry (including \text{Gaia}) measures the centre of light -- which is affected by phase illumination and differing reflectivities. Occultation astrometry is based on
the full profile of the asteroid -- irrespective of phase or reflectivity. Fig.~\ref{erosphase}  illustrates the difference between illumination and actual profile - and the consequential difference in the measured position of the asteroid. 

When a shape model is available and can be matched to the occultation chords, the centre of mass can be identified and used for reporting astrometry. The difference between centre of figure and centre of mass is illustrated in Fig.~\ref{erosshape}.

\item Once 4 or more occultations are observed over a few years and around the asteroid's orbit, the high precision of occultation astrometry renders traditional astrometry to be of little significance. Indeed, the astrometry of asteroids will even be better than \textit{Gaia}, which on any one date has high accuracy in only one direction, and measures the centre of light rather than mass.

\item For occultations observed pre-\textit{Gaia}, the astrometric position is derived using \textit{Gaia} star positions, ensuring positions are on the Gaia reference frame.

\begin{figure}
	\includegraphics[width=\columnwidth]{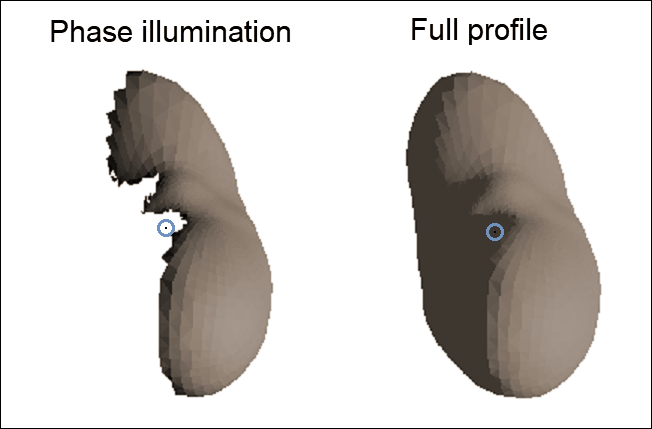}
    \caption{\textit{Illustration of the effects of solar illumination phase.} The two images show the illumination (left) and profile (right) of (433) Eros in March 2019, when Eros was at a phase angle of 51$^{\circ}$. The images are from ISAM  \protect\citep{Isam} which uses the shape model derived by  \protect\citet{ErosPhase}. The approximate location of the geometric centre of the asteroid's full profile is marked with a blue circle in both images. The geometric centre is not illuminated in the left image, with the photo-centre being displaced by about 5 per cent of the mean diameter of Eros, towards the bottom right.}
    \label{erosphase}
\end{figure}

\begin{figure}
	\includegraphics[width=\columnwidth]{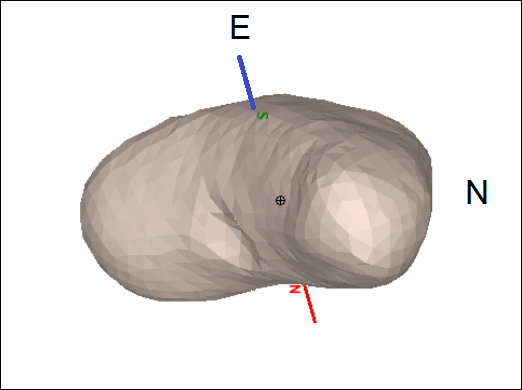}
    \caption{\textit{Illustration of the effects of irregularity.} The image shows the shape model representation of (433) Eros on 2019 Feb 25 at 5h 46m UT (with North to the right), using the same model as in Fig.~\ref{erosphase}}. The axis of rotation is shown, with its north pole oriented towards the bottom. The centre of mass is marked with a circle. At this rotational orientation, the centre of mass of the asteroid is displaced from the centre of figure by about 5 per cent of the end-to-end distance, towards the North. With the typical diameter of Eros at occultations being around 70 mas, this corresponds to a displacement of about 3 mas.
    \label{erosshape}
\end{figure}

\item A single chord observation has an across-path uncertainty set by the asteroid's diameter. As discussed in s.\ref{Uncertainties}, the across-path uncertainty is set at 40 per cent of the asteroid's diameter. As a result, the position derived from a single-chord event has an uncertainty that decreases in proportion to the diameter of the asteroid. That is, the smaller the asteroid, the greater the astrometric precision. When reliable asteroid orbits become available from \textit{Gaia} DR3, occultations will provide ongoing high-precision astrometry for many small asteroids -- with increased opportunities to study, for example, the Yarkovsky effect.
\end{itemize}

\section{Double stars}
The 4,300 asteroidal occultations prior to 2020 have resulted in the discovery of seventy-six new double stars. They have separations in the range of 0.2 to 380 mas, with sixty-one having a separation less than 50 mas.

Double star events and asteroidal satellite events are distinguishable on the following bases:
\begin{itemize}
    \item[a.]  the light curve exhibits a stepped drop at one or both the disappearance and reappearance -- caused by the sequential occultation of the two components of the star.  An occultation by a satellite does not result in a stepped light curve.
    \item[b.] the light drop is significantly less than the expected light drop -- caused by only one component of a double star being occulted. An occultation by a satellite will have the full expected light drop.
    \item[c.] the light curve has a sequence of two disappearances and reappearances. The light drops for both events will be the same for a sequential occultation of a single star by the parent body and its satellite, whilst the light drop will be different for the occultations of the components of a double star by a single body.
\end{itemize}
A further complication is the possibility of a double star being occulted by a binary asteroid. The occultation by (90) Antiope on 2015 Apr 2 is an example of this situation.

The double star solution from an asteroidal occultation may not be unique. There are four scenarios:
\begin{itemize}
    \item[a.] there are sufficient observed chords of each component to uniquely locate their paths across the asteroid. This provides a unique solution for the double star parameters.
    
    \item[b.] there are sufficient chords of one component to uniquely locate its path across the asteroid, with insufficient chords to uniquely locate the path of the other component. This leads to two solutions for the double star parameters.
 
    \item[c.] there are insufficient chords to locate the paths of both components across the asteroid. Most frequently this situation arises when there is only one observer. In this situation the asteroid is treated as circular having its assumed diameter, with there being 4 combinations of the component paths across the asteroid. This leads to four solutions for the double star parameters. 
    
    \item[d.]  Only one component of a double star is occulted. The double star parameters cannot be determined in  this situation.
\end{itemize}

The data-set includes all solutions for position angle and separation of double stars occultations. Table~\ref{doubstars} lists those stars observed prior to 2020, many of which have been published in the Journal of Double Star Observations. Where there are multiple solutions, only one solution is listed in this table.

\begingroup
    \courier
    \scriptsize
    \onecolumn
    \begin{longtable}[h]{|r|r|l|c|rr|rr|c|rl|}
 
         \caption{\hspace{0.3cm}76 double star discoveries\label{doubstars}}\\
         \multicolumn{11}{c}{Double stars discovered in an occultation before 2020, together with their parameters.}\\
         \multicolumn{11}{c}{The column $\sharp$ is the number of solutions. Only one solution is listed.}\\
         \multicolumn{11}{l}{}\\

          \hline
          Right Asc. & Declination& && Sep& $\pm$ &PA& $\pm$&&Asteroid& \\
          h\hspace{0.1cm} m\hspace{0.1cm} s\hspace{0.2cm}  & \hspace{0.1cm} $^{\circ}$\hspace{0.15cm} ’\hspace{0.05cm} ”
          \hspace{0.cm} &Star number & \# &mas &mas & $^{\circ}$& $^{\circ}$&Date&Number&Name\\
         \hline
         \endfirsthead
        
         \caption{Double star discoveries (continued...)}\\
         \hline
         Right Asc&Declination&Star number&\# &Sep&$\pm$ &PA&$\pm$ &Date&Number&Name\\
         \hline
         \endhead
         
         \hline

         \endlastfoot
        00 24 38.6&-15 28 42&UCAC4 373-000461    &1&  16.0& 0.2& 22.0& 0.7  &2017 Nov 26&     404&Arsinoe\\
        01 46 51.5&+08 16 08&Tycho2 0622-00932-1 &1&  18.7& 1.6&248.3& 3.5  &2005 Nov 11&     116& Sirona\\
        02 01 29.8&+28 31 56&UCAC4 593-004965    &1&  16.0& 7.3&139.7& 3.9  &2018 Nov 13&     476& Hedwig\\ 
        02 33 42.8&+34 20 29&Tycho2 2332-01054-1 &1&   1.1& 0.5&219.1&38.\phantom{3}  &2010 Aug 31&     695& Bella\\       
        02 44 07.3&+40 54 29&Tycho2 2849-00430-1 &1&   0.7& 0.4& 97.0&33.\phantom{5}  &2012 Jan 13&     759& Vinifera\\
        03 39 55.7&+31 55 33&HIP 17113           &4&   6.1& 2.9&148.0&21.\phantom{5}  &2017 Mar 25&     925& Alphonsina\\
        03 53 33.3&+27 18 06&Tycho2 1808-00641-1 &1&   7.8&    &275.7&      &2008 Jan 14&    1258& Sicilia\\
        05 31 33.8&+35 01 46&Tycho2 2411-01645-1 &1&   6.3& 0.9&248.7& 8.7  &1996 Nov 25&      93& Minerva\\
        05 36 29.3&+06 50 02&UCAC4 485-013038    &1&   3.5& 1.9& 37.1&26.\phantom{0}  &2012 Mar 11&      57& Mnemosyne\\
        05 40 15.8&+24 57 55&UCAC4 575-018293    &2&  23.0&    & 99.0&      &2012 Feb 18&     177& Irma\\
        05 45 34.9&+32 49 15&UCAC4 615-026159    &1&   4.5& 2.9&351.2&42.\phantom{8}  &2017 Apr 15&     187& Lamberta\\
        05 45 38.5&+18 55 02&Tycho2 1307-01531-1 &4&  36.0&    &230.8&21.\phantom{4}  &2016 Oct 18&      72& Feronia\\
        05 52 44.4&+40 10 45&Tycho2 2916-02502-1 &1&  65.4& 2.9& 71.9& 4.5  &1999 Sep 21&     375& Ursula   \\
        06 10 39.5&+21 48 08&Tycho2 1326-01111-1 &1&   4.6& 1.2& 11.7& 6.0  &2019 Oct 17&      86& Semele     \\
        06 11 42.1&+26 53 05&UCAC4 585-026030    &4&  23.0& 6.7&227.0&11.9  &2010 Oct 20&     675& Ludmilla  \\
        06 25 32.9&+23 19 38&HIP 30570           &1&  24.2& 7.6&343.3& 5.7  &2001 Sep 07&       9& Metis     \\
        06 31 17.5&+22 47 51&UCAC4 564-028451    &1& 259.9& 2.5&239.5& 0.8  &2006 Sep 19&     144& Vibilia \\
        06 37 42.7&+16 23 58&HIP 31681           &1&  63.3& 1.3&126.0& 6.5  &1991 Jan 13&     381& Myrrha \\
        06 37 51.5&+06 41 00&HIP 31694           &1&  45.9& 3.6& 27.3& 2.5  &1978 Dec 11&      18& Melpomene\\
        06 47 13.6&+16 55 35&HIP 32525           &1&   9.7& 2.2& 70.9&17.0  &2007 Mar 28&      72& Feronia\\
        06 47 28.7&+02 03 01&Tycho2 0152-00753-1 &2&  33.4& 0.1& 21.0& 0.2  &2018 Jan 29&     392& Wilhelmina \\
        06 50 29.6&+31 55 48&UCAC4 610-036507    &1&  28.1& 0.7&106.8& 4.8  &2012 Dec 03&     388& Charybdis\\
        06 52 38.5&+11 24 19&Tycho2 0755-02073-1 &1&  23.6& 0.4&101.7& 1.0  &2018 Feb 09&     191& Kolga\\
        06 53 06.5&+22 04 15&Tycho2 1343-01414-1 &2&  34.0&    &275.0&      &2011 Apr 01&     554& Peraga\\
        07 00 39.3&+12 44 24&HIP 33753           &1&   5.9& 1.0& 54.9&10.4  &2018 Dec 10&     479& Caprera\\
        07 08 31.0&+21 59 10&HIP 34452           &2&  39.0&    & 88.9& 0.1  &1974 Aug 29&Saturn V& Rhea\\
        07 14 41.2&+29 52 39&Tycho2 1908-00844-1 &2& 170.4& 0.6&303.0& 0.1  &2006 Nov 29&     578& Happelia\\
        07 21 07.2&+43 16 00&UCAC4 667-050740    &1&  54.2& 7.2&238.6& 7.9  &2013 Jan 27&     536& Merapi\\
        07 27 09.1&+11 57 18&HIP 36189           &1&  13.0& 0.7&231.9& 4.0  &2003 Mar 23&     704& Interamnia\\
        07 37 31.7&-00 59 30&Tycho2 4831-00302-1 &2&  28.0&    &  3.0&      &2009 Dec 01&     130& Elektra\\
        07 44 05.6&-04 15 32&UCAC4 429-038378    &1&   5.3& 3.8&339.8&44.\phantom{3}  &2017 Mar 23&     978& Aidamina\\
        07 52 45.6&+18 49 37&HIP 38465           &1&   7.8& 2.9&103.0&      &2002 Dec 24&     334& Chicago\\
        08 20 02.0&-02 35 39&UCAC4 438-046254    &1&  42.0&    &351.0&      &2018 Jan 13&      57& Mnemosyne\\
        08 25 01.7&+28 33 55&Tycho2 1947-00293-1 &3& 117.5&    &159.5&10.9  &2006 Dec 18&      87& Sylvia\\
        08 25 36.3&+28 08 29&Tycho2 1947-00290-1 &4&  18.5&    & 82.2&41.\phantom{9}  &2019 Dec 12 &    87 &Sylvia\\
        08 34 20.9&+10 12 31&UCAC4 502-048751    &2&   2.6&    &283.6&      &2014 Jan 18&     664& Judith\\
        09 01 30.7&+24 27 28&UCAC4 573-045670    &1&  27.3& 0.7& 11.4& 1.5  &2013 Feb 06&      92& Undina\\
        09 05 39.3&+22 34 53&UCAC4 563-047083    &2& 154.0& 3.3&102.5& 1.2  &2013 Dec 28&     141& Lumen\\
        09 11 41.9&-01 06 01&UCAC4 445-049380    &1&  10.0&    & 57.0&      &2016 Mar 19&     695& Bella\\
        09 16 47.1&+08 14 18&Tycho2 0819-00852-1 &1&   7.4& 1.8&126.1&15.0  &2009 Apr 16&     336& Lacadiera\\
        09 25 45.2&+20 22 50&HIP 46249           &1&   4.4& 2.2& 45.2&31.\phantom{7}  &2011 Jan 24&     160& Una\\
        10 07 45.1&+06 44 36&Tycho2 0250-00557-1 &1&   2.4& 3.2&128.1&      &2009 Nov 09&      79& Eurynome\\
        11 27 32.4&-09 32 02&UCAC4 403-053751    &1&  99.1& 1.5&359.7& 5.5  &2006 Mar 02&      12& Victoria\\
        11 46 48.4&+05 52 31&Tycho2 0278-00748-1 &1&   0.2&    & 27.9&      &2018 May 22&     201& Penelope\\
        12 01 10.1&+02 58 27&Tycho2 0283-00694-1 &2&  37.3& 1.4&124.6& 3.6  &2015 Apr 02&      90& Antiope\\
        12 17 48.0&+03 56 50&UCAC4 470-046569    &1&   4.0& 4.6&302.0&36.\phantom{5}  &2018 Mar 13&     784& Pickeringia\\
        12 28 34.1&+03 51 38&UCAC4 470-046799    &2&  18.3& 1.0& 28.8& 1.8  &2019 Mar 19&    1072& Malva\\
        13 30 52.2&-05 06 05&Tycho2 4972-00102-1 &4&  48.2& 6.3& 20.5& 4.0  &2018 May 06&      34& Circe\\
        13 37 15.8&+04 25 03&HIP 66446           &1&  25.1& 5.4&275.6& 8.5  &2001 Mar 15&     423& Diotima\\
        14 19 06.6&-13 22 16&HIP 69974           &4&   8.0& 0.1& 13.0& 1.3  &2006 Apr 12&     305& Gordonia\\
        14 41 46.0&-16 36 09&Tycho2 6154-00401-1 &2&   7.9& 2.2& 89.8&22.\phantom{5}  &2003 Apr 21&     210& Isabella\\
        15 35 06.2&-25 06 20&HIP 76293           &2&  39.5&    & 80.2&      &2007 May 18&    1177& Gonnessia\\
        15 46 28.9&-09 10 35&Tycho2 5597-01223-1 &1&  49.8& 2.5&333.1& 4.0  &2016 Aug 05&     511& Davida\\
        15 46 55.9&-12 05 02&UCAC4 390-065591    &4&  15.0& 7.5&183.8&28.\phantom{6}  &2015 Feb 12&     107& Camilla\\
        15 54 10.2&-09 44 50&Tycho2 5614-00026-1 &1&  28.0&    &198.0&      &2002 May 10&     638& Moira\\
        16 05 26.6&-19 48 06&HIP 78821           &1&  98.0& 5.8&311.3& 4.4  &1971 May 14&   Jupiter I& Io\\      
        16 19 32.2&-40 03 30&UCAC4 250-090193    &1&   3.2& 0.4&194.7&11.3  &2019 Aug 03&    4004& List'ev\\
        16 31 06.1&-14 55 20&UCAC4 376-077172    &1&   5.3& 1.2&132.9&12.9  &2019 Apr 22&     145& Adeona\\
        17 00 36.9&-17 13 46&Tycho2 6223-00442-1 &4&  84.8& 1.8&266.1& 1.2  &2012 Aug 12&      52& Europa\\
        17 10 51.6&-21 18 20&UCAC4 344-090563    &1&  47.0&    &145.0&      &2015 Aug 05&      92& Undina\\
        17 56 19.4&-19 01 43&UCAC4 355-118771    &1&   1.4& 3.3&247.0&      &2013 Jul 24&    1271& Isergina\\
        18 18 05.9&-12 14 33& HIP 89681          &2&  29.6& 0.4& 64.2& 0.8  &2019 Sep 20&    713& Luscinia\\
        18 18 39.7&-20 02 46&Tycho2 6273-01033-1 &4&  42.5&    &146.5&23.\phantom{6}  &2012 Mar 20&      44& Nysa\\
        19 16 13.0&+21 23 26&HIP 94703           &1&   2.6& 0.7&283.7&20.\phantom{2}  &1983 May 29&       2& Pallas \\
        19 30 57.4&-22 36 13&UCAC4 337-189531    &1& 193.9& 4.2& 54.0& 3.8  &2019 Jul 23&    3130& Hillary\\
        19 33 16.3&-04 17 33&UCAC4 429-099842    &4& 379.2& 1.6&236.1& 0.5  &2013 Aug 15&     611& Valeria\\
        20 03 37.2&-33 15 16&Tycho2 7444-01434-1 &4&  28.5& 0.5&283.0& 1.7  &2013 Aug 15&     481& Emita\\
        20 30 30.5&+01 24 49&UCAC4 458-117279    &1&  30.4&    &115.6& 7.5  &2015 May 12&     849& Ara\\
        20 45 33.8&-04 26 20&Tycho2 5186-00724-1 &1&   1.7& 0.2&272.8& 5.8  &2009 Jul 28&     732& Tjilaki\\
        21 09 41.9&-12 37 55&HIP 104465          &1&   9.2& 4.8&171.5& 6.8  &2012 Nov 25&     168& Sibylla\\
        21 12 25.8&-11 49 07&Tycho2 5780-00308-1 &1&  14.3& 0.4& 74.2& 2.7  &2017 Aug 23&     834& Burnhamia\\
        21 15 05.9&-03 06 19&UCAC4 435-115475    &4& 368.2& 6.6& 94.5& 2.9  &2018 Nov 09&     409& Aspasia \\
        22 03 49.9&-02 48 13&UCAC4 436-117479    &1&   1.0&    & 38.0&      &2014 Dec 09&    3950& Yoshida \\
        23 36 23.2&+02 06 08&HIP 116495          &1&  11.9& 2.7&127.4&12.9  &2006 May 05&       7& Iris\\
        23 38 27.1&+02 15 02&HIP 116660          &1&  17.8&    & 19.0&      &2009 Sep 09&    1157& Arabia\\
        23 54 28.2&+25 32 46&UCAC4 578-136317    &1& 144.8& 1.4&192.4& 0.7  &2009 Jul 19&     790& Pretoria\\

    \end{longtable}
\endgroup

\twocolumn
\section{Binary asteroids \& Asteroid satellites}
Twelve events in V3.0 of the data-set involve an occultation by a known satellite of an asteroid -- with one event involving two satellites of the asteroid. Another three events involve Pluto \& Charon. These are listed in Table~\ref{Knownsatellites}, which shows that occultation observations provide accurate separation and position angles, and frequently provide measurement of the size and shape of the satellite. When there are multiple events involving an asteroid system [such as with (90) Antiope] it becomes possible to refine the orbital parameters of the system, especially when the centre of mass of the main body can be determined from a fit of the main body to a shape model.

\begin{table*}
	\caption{List of occultations by known satellites of asteroids. The adjacent columns of Sepn and PA give the separation and position angle of the satellite from the main body. The adjacent columns of Major, Minor and PA give the size and orientation of an ellipse fitted to the occultation chords for the satellite. The quality of that fit is indicated in the final column. For the event involving Kleopatra, the satellite was either Alexhelos or Cleoselene.}
	\label{Knownsatellites}
    \begingroup
        \courier
        \small	
        \tabcolsep=0.10cm
        \begin{tabular}{rlllrrrrrl}
            \hline Asteroid & Asteroid& Satellite & Date& Sepn & PA & Major & Minor&PA&Quality of fit \\
            Number&Name &Identification & &mas & deg & km & km&deg&to the satellite\\
            \hline
            22&Kalliope &(22) 1 Linus&2006 Nov 7&246.0&318&33&26&4&Size \& shape\\
            87&Sylvia&(87) 1 Romulus&2013 Jan 6&599.9&100&33&18&122&Size \& shape\\
            87&Sylvia&(87) 1 Romulus&2019 Oct 29&513.0&74.9&36&20&291&Size \& shape\\            "&\phantom{Syl}"&(87) II Remus&\phantom{20}"\phantom{ Oc}"\phantom{
            2}"&264.5&90.7&12&9&95&Size \& shape\\            90&Antiope&S/2000 (90) 1&2008 Jan 2 &70.0&144&95&81&340&Location only\\
            90& Antiope &S/2000 (90) 1&  2011 Jul 19 & 125.0& 192& 102& 89&171&Size \& shape\\
            90& Antiope & S/2000 (90) 1&2015 Apr 2 & 89.3&333& 94& 81&316&Size \& shape\\
            90& Antiope & S/2000 (90) 1&2020 Mar 15& &&&&&Relies on shape model\\
            93& Minerva&(93) II Gorgoneion&2014 Sep 6 &164.8& 230& 7&7&&Approximate size\\
            216&Kleopatra&(216) I Alexhelos ? & 1980 Oct 10&525&149&9&9&&Location only\\
            617& Patroclus &(617) 1 Menoetius&2010 Jun 14& & &68&68&&Main body not seen\\
            617&Patroclus&(617) 1 Menoetius&2013 Oct 21& 247.0&266& 93& 117& 354&Size \& shape\\
            90482 & Orcus & 90482 (I) &  2017 Mar  7 & & &295&295&&Main body not seen\\
            & Pluto & Charon &2005 July 11& & &1212&1212&&Main body not seen\\
            & Pluto & Charon &2008 Jun 22& 572.0&70&1212&1212&&Location only\\
            &Pluto &Charon&2011 Jun 23& 667.0&267& 1200&1200& & Location only\\
    	\end{tabular}
	\endgroup
\end{table*}

Sixteen events (listed in Table~\ref{satellites}) in V3.0 of the data-set involve observations (all being single-chord observations) which \textit{might} be explained by the presence of an unknown satellite. These events must be treated with considerable caution. Early visual observations reported many ‘false' occultations, most likely associated with observer fatigue or poor seeing. It follows that visual claims to a satellite discovery must be treated with similar caution. Electronic recording of an occultation should be reliable - provided there is adequate signal-to-noise, and the event duration is not very short. Table~\ref{satellites} identifies those observations made visually, and those electronic observations reported as uncertain. It also provides the magnitude drop, the duration of the event, and the corresponding minimum diameter. The light-curve for the event involving (18) Melpomene \citep{Melpomene..1980AJ.....85..174W} appears definite, but searches using a 2.2m telescope \citep{1988LPI....19..405G} and the Hubble Space Telescope \citep{1999Icar..137..260S} failed to find it.

As set out in section \ref{reductions}, the reference time for events involving satellites is set as the mean of the times involving the satellite. When available, the motion of the satellite on the fundamental plane is adjusted using the Miriade\footnote{\textcolor{blue}{http://vo.imcce.fr/webservices/miriade/}} system of Paris Observatory, to more accurately bring multiple events for the satellite together.

The data-set includes the separation and position angle of the satellite relative to the primary body. For three events -- Pluto (Charon) on 2005 Jul 11, (617) Patroclus on 2010 June 14, and (90482) Orcus (Vanth) on 2017 Mar 7 \citep{Sickafoose_2019} -- an occultation by the satellite was observed, but not the main body. For satellites of the major planets (such as Charon), the reduction is based on the known position of the satellite relative to the planet -- with no separation and position angle being derived. For asteroids, an ‘artificial' event for the main body is included using special event codes. The astrometric position of the satellite is the offset of the satellite from the artificial body combined with the offset of the artificial body from the star. The separation and position angle  of the satellite cannot be derived.

The information provided in the data-set for a satellite varies from mere detection through to full details of its dimensions. 

\begin{table}
	\centering
	\caption{List of events where the  observation of one observer \textit{might} be explained by the presence of a satellite. In all cases there were no confirming observations. Sepn and PA are the separation and position angle of the putative satellite from the main body. Mag drop is the expected drop in magnitude for the occultation, Drn Sec is the reported duration of the occultation, and Size km is the corresponding diameter of the satellite. The last column flags events observed visually (V), and those recorded electronically that were reported as uncertain (?)}
	\label{satellites}
    \begingroup
        \courier
        \small
        \tabcolsep=0.07cm
    	\begin{tabular}{rllrrrrrr}
            \hline & Asteroid   & Date&  Sepn & PA&Mag&Drn &   Size &C \\
            No.&Name & & mas & deg &drop &Sec& km& \\
            \hline
              6 & Hebe & 1977 Mar  5& 531 & 336      &6.8 &0.5  &20&V \\
             15 & Eunomia  & 2019 Apr 27& 316 & 251  &0.1 &2.8 &80&? \\
             18&Melpomene&1978 Dec 11&910&223        &1.3&5.8   &49&  \\
             71&  Niobe &  2005 Feb 10&  275 & 120   &4.5 &1.0  &10&V \\
             96& Aegle & 2002 Aug 10&  240 & 92      &2.7&2.2   &43&V \\
             98& Ianthe& 2004 May 16&  1177   & 289  &3.4 &0.11 & 4&  \\
             146& Lucina & 1982 Apr 18 & 1351 & 60   &3.2&0.6   &10&  \\
             412& Elisabetha & 2016 Mar 17& 211&309  &1.8&0.2   & 4&V \\
             532& Herculina& 1978 Jun  7& 831 & 223  &4.5&5     &45&  \\
             578& Happelia & 2017 Feb 24 & 37 & 292  &2.1&0.17  & 3&? \\
             595& Polyxena   & 2008 Feb  3 &152&338  &1.3&1.7   &23&V \\
             776& Berbericia  & 2011 Aug  5 & 132&19 &2.0&1.0   &11&V \\
             911& Agamemnon  & 2012 Jan 19  & 94&93  &7.4&0.18  & 9&  \\
            2258& Viipuri  & 2013 Aug  3 & 89 & 256  &8.0&0.55  & 3&? \\  
            2258& Viipuri  & 2018 Sep 19& 86 & 241   &4.1&0.7   &10&  \\
            2494& Inge  & 2016 Nov  7& 36 & 249      &2.3 &0.6  & 8&  \\
    	\end{tabular}
	\endgroup
\end{table}

\begin{figure}
	\includegraphics[width=\columnwidth]{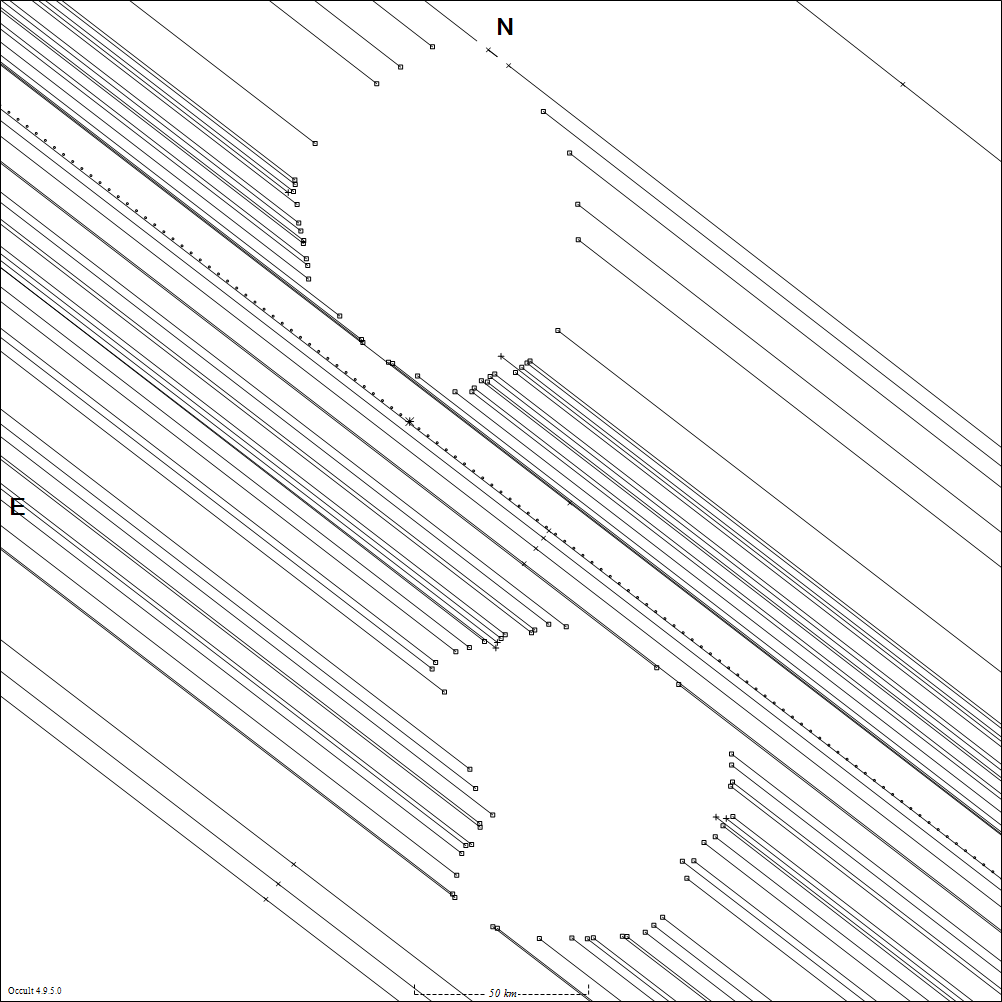}
    \caption{Occultation by the binary asteroid (90) Antiope observed in the USA on 2011 July 19. The resolution is of the order of 1~km.}
    \label{fig1}
\end{figure}

\begin{figure}
	\includegraphics[width=\columnwidth]{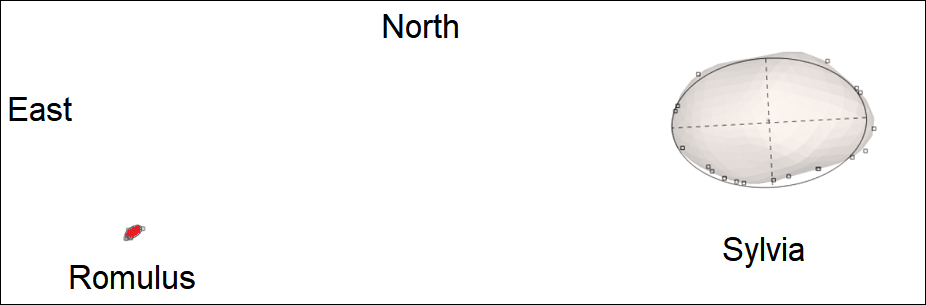}
    \caption{Occultation by the asteroid (87) Sylvia and its satellite Romulus, observed in Europe on 2013 Jan 6. A shape model has been fitted to the main body of Sylvia, giving a \textit{volume}-equivalent diameter of about 276~km. Two occultation events separately establish the diameter of Romulus as being about 25~km; it is well resolved in this occultation, as is its position with respect to the main body.}
    \label{figSylviaRomulus}
\end{figure}
\subsection{Resolution attainable}
Fig.~\ref{fig1} shows the binary asteroid Antiope. The resolution is of the order of 1km, vastly better than images\footnote{\textcolor{blue}{https://www.boulder.swri.edu/merline/press/press2/fig1.html}} by the Southwest Research Institute using adaptive optics on the 10m Keck telescope. 

Fig \ref{figSylviaRomulus} is of Sylvia, an asteroid which has two satellites (Romulus and Remus) of much smaller size. As clearly evident, the occultation observation well-defines the position of Romulus.

\citet{Sylvia} reported a diameter of $20 \pm 5$ km using $\Delta$\textsubscript{\textit{m}} between the main body and satellite. \cite{2012AJ....144...70F} derived the mean diameters of Romulus as 5--16 km and Remus as 9--12 km, by assuming the satellites had the same density as the main body of Sylvia (see para 7.2 of that paper). A greatly different diameter for Romulus was derived by \cite{2016Icar..276..107D}, with a prolate spheroid of 41 $\pm$27 x 30 $\pm$16~km, and an alternative possibility of 82 $\pm$7 x 21 $\pm$2~km. 

\begin{figure}
	\includegraphics[width=\columnwidth]{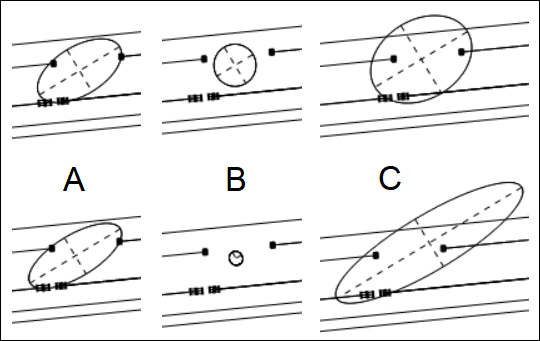}
   \caption{Detail of the occultation by Romulus on 2013 Jan 6. The ‘A' images show the maximum and minimum ellipses that fit the occultation chords. The ‘B' images show the upper and lower diameter limits from \protect\cite{2012AJ....144...70F}. The ‘C' images show the two preferred possibilities from \protect\cite{2016Icar..276..107D} -- without indicating the uncertainty range for each.}
    \label{figRomulus2013}
\end{figure}

Romulus has been observed in two occultations -- 2013 Jan 6, and 2019 Oct 29 \citep{SylviaTriple}.
The 2013 occultation recorded two well-defined chords, and a ‘Miss' chord that tightly constrained the size of Romulus. The major axis of the ellipse fitted to the occultation chords is between 40 \& 16~km. The minor axis is correspondingly between 18 \& 35~km. (Variations in the length of the axes are not independent; a decrease in one axis alters the orientation of the fitted ellipse which in turn increases the length of the other axis in order to fit the chords -- and \textit{vice versa}. The extrema are shown by group A in Fig.\ref{figRomulus2013}.) After allowing for the interrelationship between the variations of the axes of the ellipse, the mean diameter of Romulus is 25 $\pm$1~km. [\cite{BERTHIER2014118} derived a mean diameter of 23.1 $\pm$0.7~km from this occultation.] Fig. \ref{figRomulus2013} shows the observed occultation chords for the 2013 occultation, the ellipse range from \cite{2012AJ....144...70F} and the two preferred solutions from \cite{2016Icar..276..107D}. 

\begin{figure}
	\includegraphics[width=\columnwidth]{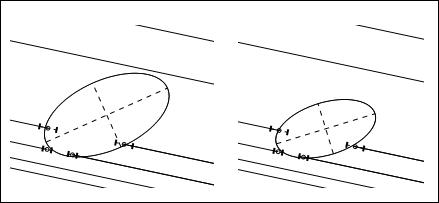}
	\caption{The 2019 October 29 occultation by Romulus, which is less well-constrained than the 2013 occultation. The two images illustrate the maximum and minimum ellipses that fit the occultation chords. This event occurred after version V3.0 of the data-set.}
    \label{figRomulus2019}
\end{figure}

The 2019 occultation by Romulus was less well constrained. Fig. \ref{figRomulus2019} shows the two limiting ellipses that fit the chords of that event. The major axis of the fitted ellipse is between 29 \& 15~km; the minor axis is correspondingly between 20 \& 38 km - giving a mean diameter of 24 $\pm$4~km. 

These two occultation measures of the diameter of Romulus are in full agreement, giving confidence that the mean diameter of Romulus is about 25 $\pm$1~km. The diameter found in \cite{2012AJ....144...70F} by assuming the density of Romulus was the same as for Sylvia is inconsistent with the occultation result.

\section{Shape models and occultations} \label{ShapeModels}
  The great majority of shape models are derived from light-curve inversion techniques. Shape models were initially limited to convex solutions, but now concave solutions are being derived. 
  The number of asteroids having one or more shape models has greatly increased in recent years. Additionally occultations have been used to refine shape models using tools such as ADAM \citep{ADAM}, with a recent example being of (16) Psyche \citep{Psyche}.

In version 3.0 of this data-set we have identified each shape models available from two sources:
\begin{itemize}
    \item DAMIT\footnote{\textcolor{blue}{https://astro.troja.mff.cuni.cz/projects/damit}} \citep{Damit} 
    \item ISAM\footnote{\textcolor{blue}{http://isam.astro.amu.edu.pl/}} \citep{Isam}
\end{itemize}
While there is a high degree of commonality of shape models between these sources, there are instances of models being present in one but not the other.

\begin{figure}
	\includegraphics[width=\columnwidth]{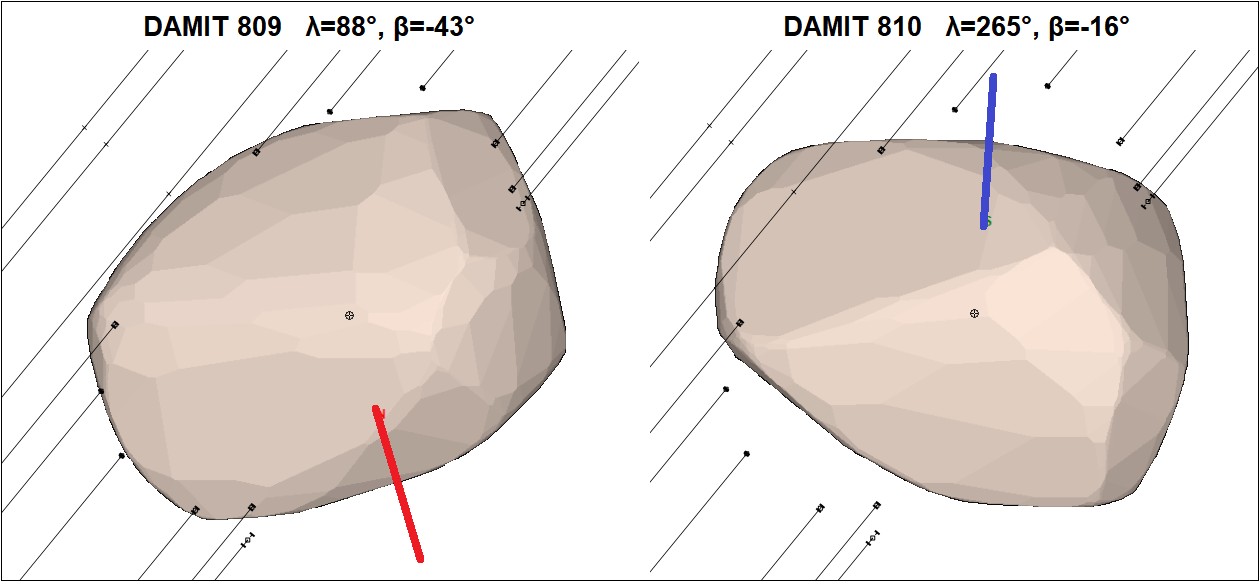}
    \caption{The fit of the chords from the 2013 Aug 24 occultation by (489) Comacina to two different shape models from \protect\cite{Damit809-810}. The respective ecliptic longitude and latitude of the pole are shown above each plot. The relative motion of the star is from left to right. The polar axes visible on the images are the north (left) and south (right) poles. The occultation chords closely match the shape model on the left, but not that on the right - indicating that the left shape model has the correct axis of rotation.}
    \label{Comacina}
\end{figure}

We have only derived diameters using the shape models when justified. At this time the fits are all ‘visual’ fits; we have not attempted a mathematical fit of occultation chords to the shape models. There are seven categories (0 to 6) of fits: 
\begin{itemize}
    \item[] \textbf{0}  No fit was made, there being too few chords to make a fit to the shape model.
\item[] \textbf{1} Bad occultation data. Occasionally it is evident that the occultation observations are inconsistent in some manner, such that no sensible fit to a shape model is possible.

\item[] \textbf{2}  Model wrong. Shape models have two main sources of error. Firstly, frequently there is an ambiguity in the solution for the axis of rotation, leading to different shape models. Occultation observations can resolve this when the occultation chords match one model but not the other, as illustrated in fig \ref{Comacina}.

Secondly, the light curve inversion process is complex. Sometimes none of the shape models match the occultation chords -- indicating inadequacies in the shape model derivation.

\item[] \textbf{3}  Minimum diameter.  This is usually associated with a single-chord event having a length greater than about 80 per cent of the asteroid's assumed diameter. For these events the chord is aligned with the greatest along-chord dimension of the shape model to derive a minimum diameter consistent with that single chord.

\item[] \textbf{4}  Diameter but no fit. There are a number of events where the match of the observed chords to the shape model is very approximate. In such situations it may be possible to estimate an approximate diameter using that shape model. Clearly such estimates of diameter must be treated with caution. 

\item[] \textbf{5} and \textbf{6} The occultation chords provide either a poor (5) or good (6) fit to the shape model. The categorization is a subjective assessment based on the visual matching of the chords to the shape model. There are no events where the matching is ‘perfect'. This inevitably leads to a degree of uncertainty in the diameter of the asteroid, and possibly in the centre of mass astrometry derived from the occultation. Nevertheless, we believe these events provide reliable measures of the diameter of asteroids, and accurate centre of mass astrometry.
\end{itemize}

\subsection{Shape models -- diameters and astrometry}

\cite{Inversion} combined shape models for 44 asteroids with occultation observations to derive \textit{volume}-equivalent diameters for those asteroids. In what follows we will explore the issues associated with matching shape models to occultations in more detail, as well as measuring diameters from occultations where no shape model is available.

For an irregular object like an asteroid, there are two types of mean diameter. Firstly, the \textit{volume}-equivalent diameter -- which is the diameter of a sphere having the same volume as the asteroid. This is relevant for assessing the density of an asteroid when its mass is known -- as in \cite{2017A&A...601A.114H}. Secondly, the \textit{surface}-equivalent diameter -- which is the diameter of a sphere having the same surface area. This is relevant to radiation emission or light reflection. Assessment of the models in DAMIT shows that on average the \textit{surface}-equivalent diameter is 5.8 per cent greater than the \textit{volume}-equivalent diameter. However for 10 asteroids it is more that 20 per cent larger, and for 310 asteroids it is more than 10 per cent larger -- making it important to specify whether a diameter is \textit{volume}-equivalent or \textit{surface}-equivalent diameter. In this paper we are primarily concerned with \textit{volume}-equivalent diameters. We refer to a mean diameter when the distinction is not material or relevant.

Shape models are defined by an array of vectors originating at the centre of volume of the asteroid, normalised to the longest vector. The \textit{volume}-equivalent and \textit{surface}-equivalent radii, expressed as a fraction of that longest vector, are computed in accordance with \cite{DOBROVOLSKIS1996698}. Changing the scale of the shape model as plotted on the fundamental plane correspondingly changes the length of that longest vector in the coordinate units of the fundamental plane. When the scale of the shape model is adjusted to match the occultation events on that plane, and the coordinate units are km, the radius of the asteroid in km is simply the relevant fraction of the length of that longest vector (with the corresponding diameters being twice those radii).

The visual fit of the occultation chords to a shape model is inherently subjective. There are very few instances where the occultation chords have a perfect match with a shape model profile. Causes include errors in the occultation data, and deficiencies in the shape model. The fitting process involves making a visual best fit of the shape model to all the plotted error bars of the chords. Two fits are made. One is for a best fit to one extrema of the error bars; the other to the other extrema. The result is a measurement of maximum and minimum diameters consistent with the shape model. When a small number of observed chords have large uncertainty bars compared to the other chords, they will be discounted in the visual fitting process. When all the chords have relatively large error bars, any determination of diameter would be meaningless and the chords are not fitted to the shape model.

\begin{figure}
	\includegraphics[width=\columnwidth]{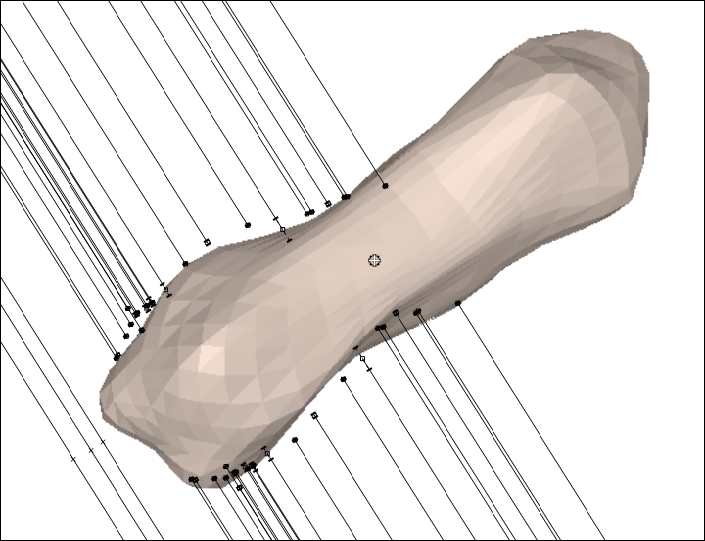}
	\caption{The 2015 March 12 occultation by (216) Kleopatra, plotted against the shape model from \protect\cite{2017A&A...601A.114H}. The centre of volume/mass is marked by a circle. The centre of a best-fit ellipse to the observations is displaced from the centre of volume/mass of the asteroid by about 48 km, equal to 38 per cent of its mean diameter. The corresponding angular displacement for this event is 30 mas.}
    \label{figKleopatra}
\end{figure}

The centre of volume (and of mass, assuming uniform density) is at the origin of the vectors. Its location is readily identifiable. When the shape model is adjusted to match the occultation chords, the astrometry can be based on the location of the centre of mass of the asteroid, rather than the centre of figure of the occultation chords. Fig \ref{figKleopatra} is an example which shows an occultation by (216) Kleopatra on 2015 March 12. The location of the centre of volume/mass (indicated by a circle) is well established by the fit of the occultation chords to the shape model; the centre of figure of the occultation chords is well displaced from the centre of mass. While this is an extreme example, the principle applies to all occultations.

Shape models continue to grow in number, as well as revised, on the basis of new data (both photometric and occultation results). Accordingly shape model matching in the data-set is expected to improve in future versions.

\section{Diameters}
Since its inception, the primary objective of observing asteroidal occultations has been to determine asteroid diameters. The reason for doing this is well explained by \cite{VERES201534} as:
\begin{quotation}Asteroid diameters are critical to understanding their dynamical and morphological evolution, potential as spacecraft targets, impact threat, and much more, yet most asteroid diameters are uncertain by > 50 per cent because of the difficulties involved in calculating diameter from apparent brightness. The problem is that an asteroid’s apparent brightness is a complicated function of the observing geometry, their irregular shapes, rotation phase, albedo, lack of atmosphere, and their rough, regolith-covered surfaces. Most of these data are unknown for most asteroids.\end{quotation}

The majority of asteroid diameters are derived from their absolute magnitude, as in \cite{VERES201534}. Three spacecraft missions (\textit{IRAS}, \textit{AKARI AcuA}, and \textit{WISE/NEOWISE} - \textit{supra}) used thermal infrared to measure asteroid diameters. All rely on calibration against a relatively small number of asteroids having diameters determined by other means -- such as speckle interferometry, radar images, and occultations. In recent times there has been some controversy about the results from \textit{NEOWISE}, with criticism \citep{2018Icar..314...64M}, counterclaim \citep{2018arXiv181101454W} and further criticism \citep{2018arXiv181206516M}. While radar observations can provide very high resolution for near-earth objects, occultations generally provide the best resolution. However the large number of occultation results are not well known by reason of their lack of publication in literature.

The data-set contains, whenever possible, diameters from fitting the occultation chords to shape models. Otherwise, the dimensions and orientation of an ellipse fitted to multiple occultation chords. However the ellipse data must not be read alone; in a number of situations one or more parameters used in the least squares solution for that ellipse need to be fixed to avoid a divergent iteration. The data-set includes flags to indicate which parameters have been set to an assumed value. 

For all single-chord events, and some multi-chord events, the distribution of the chords does not justify fitting to a shape model or ellipse. In such cases there are flags to indicate the solution was limited to a circle having the assumed diameter of the asteroid, or a circle whose diameter is determined on the basis of the chords. Such events are only relevant for astrometry. Unfortunately past versions of the data-set have not adequately differentiated between measured and assumed diameters. This has been rectified in V3.0 of the data-set.

\subsection{Verification using visiting spacecraft determinations}\label{IRverify}
An occultation observation usually provides a profile at a single orientation. (Exceptions are when observers are separated by a large distance [especially on different continents] or the asteroid's shadow is moving unusually slowly across the Earth -- where differences in the event times may change the rotational orientation of the asteroid. Note also the first point in section \ref{Future}).  Combining an occultation with a shape model should give the \textit{volume}-equivalent or \textit{surface-equivalent} diameter of the asteroid. It is appropriate to verify that this is indeed the case.

A small number of asteroids have had their \textit{volume}-equivalent diameters measured by visiting spacecraft -- diameters that are more accurate than any Earth-bound observations. Those that have also been observed in occultation with enough chords to determine a diameter are (1) Ceres, (4) Vesta, (21) Lutetia, (253) Mathilde, (433) Eros, and (486958) Arrokoth. Only Mathilde doesn't have a reliable shape model.

\subsubsection{Asteroids with shape models}\label{ModelFits}
Comparison with the satellite dimensions and those determined by occultation are:
    \begin{itemize}

    \item (1) Ceres. \label{CeresRef} The \textit{Dawn} spacecraft mission \citep{Park515} determined the dimensions of Ceres as 964.4 x 964.2 x 891.8~km, for a \textit{volume}-equivalent diameter of 939~km. (From the shape model\footnote{\textcolor{blue}{https://sbn.psi.edu/pds/resource/dawn/dwncfcshape.html}}, which includes the cratering, the \textit{surface}-equivalent diameter is 0.7 per cent larger). There are two occultations by Ceres where the chord distribution provides a reliable determination of its size:
    \begin{itemize}
        \item 1984 Nov 13, with 13 chords across the asteroid - all recorded photoelectrically. Fig \ref{Ceres} shows a fit of the chords to the \textit{Dawn} shape model. The derived \textit{volume}-equivalent diameter is 937$\pm{3}$~km, essentially the same as the \textit{Dawn} \textit{volume}-equivalent diameter. (A best-fit ellipse has dimensions of 960 $\pm{4}$ x 908 $\pm{3}$~km - giving a mean diameter of 934~km, which is 0.3 per cent smaller than the \textit{Dawn} \textit{volume}-equivalent diameter.)
    \begin{figure}
    	\includegraphics[width=\columnwidth]{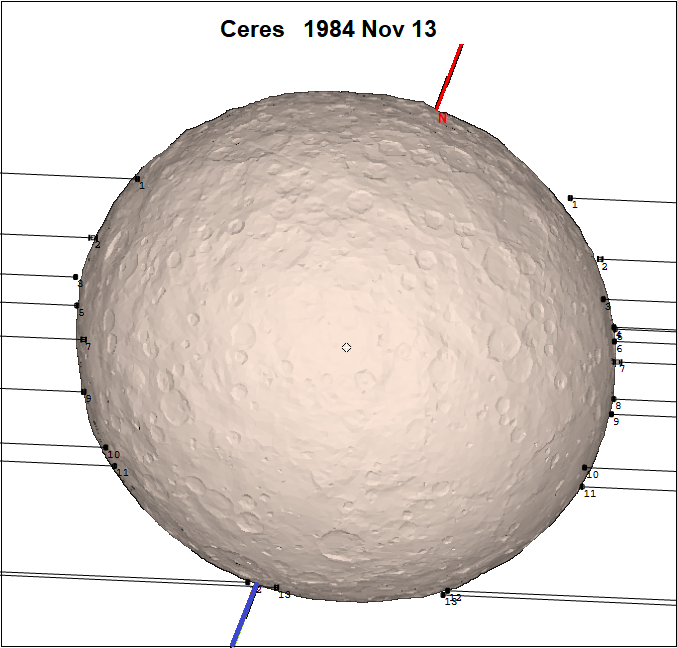}
        \caption{The fit of the chords from the 1984 Nov 13 occultation by (1) Ceres to the \textit{Dawn} shape model. The relative motion of the star is from right to left.}
        \label{Ceres}
    \end{figure}  
    \item 2013 Oct 25, with 9 chords across the asteroid. Three of the chords have relatively large uncertainties in their event times, increasing the uncertainty in the fit to the shape model. The derived \textit{volume}-equivalent diameter is 937$\pm{9}$~km, which (again) is essentially the same as the \textit{Dawn} diameter.
    
    \end{itemize}

    \item (4) Vesta. The \textit{Dawn} spacecraft mission \citep{Preusker2016} determined the dimensions of Vesta as 570.4 x 555.4 x 447.6 km, for a \textit{volume}-equivalent diameter of 521.5~km. (From the shape model\footnote{\textcolor{blue}{https://sbn.psi.edu/pds/resource/dawn/dwnvfcshape.html}}, which includes the cratering, the \textit{surface}-equivalent diameter is 2.9 per cent larger). There is only one occultation by Vesta where the chord distribution provides a reliable determination of its size -- a 2-minute occultation on 1991 Jan 4 with 19 chords across the asteroid. Many of the chords are from visual observers (which was typical in those days), and have associated uncertainties. In particular, it was not uncommon for the time of disappearance for a visual observer to be late. There are 7 video chords -- but unfortunately they are all at one side of the asteroid. Fig \ref{Vesta} shows a fit of the chords to the \textit{Dawn} shape model, with the area of high dependence on visual chords being the left 60 per cent. The derived \textit{volume}-equivalent diameter is 524~km, 0.5 per cent larger than the \textit{Dawn} mean diameter. (A best-fit ellipse has dimensions of 558 $\pm{4}$ x 458 $\pm{3}$~km -- giving a mean diameter of 506~km, which is 4 per cent smaller than the \textit{Dawn} mean diameter.)
    
    \begin{figure}
    	\includegraphics[width=\columnwidth]{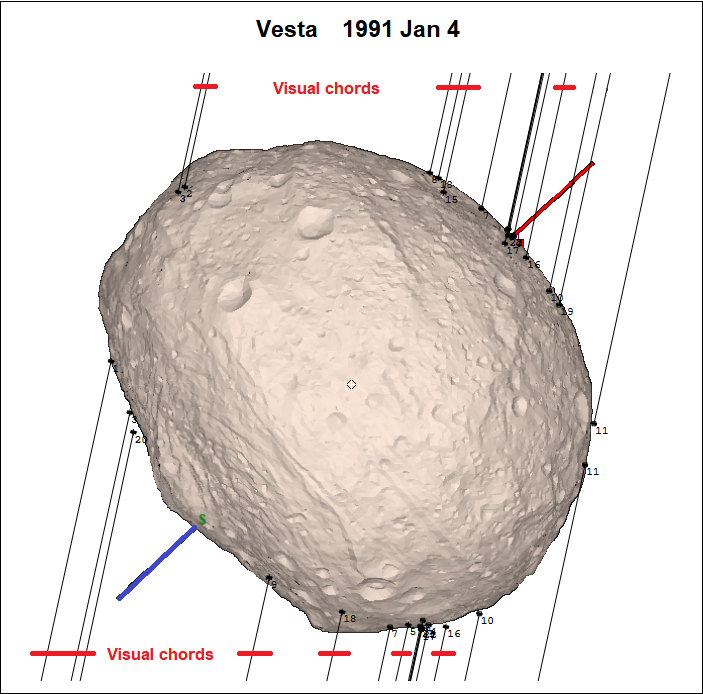}
        \caption{The fit of the chords from the 1991 Jan 4 occultation by (4) Vesta to the \textit{Dawn} shape model. The chords obtained by visual observations are marked in red. The relative motion of the star is from top to bottom.}
        \label{Vesta}
    \end{figure}

    \item (21) Lutetia. From the Rosetta spacecraft flyby in 2010, \cite{Sierks487} derived axes of 121 $\pm{1}$ x 101 $\pm{1}$ x 75 $\pm{13}$~km, giving a \textit{volume}-equivalent diameter of 98 $\pm{2}$~km. The southern hemisphere was not seen during the flyby, with the 3rd axis being estimated using a pre-flyby shape model which matched the shape of the imaged part to within 5 per cent. Several occultation observations can be matched to modern shape models; the results are shown in Table \ref{Lutetia}. Three of the 5 events (asterisked) have well-distributed chords and a good fit to shape models (2016 Sep 24, 2019 Nov 25 and 2020 Jan 26). Fig \ref{Lutitia} shows the chords of the 2019 Nov 25 event fitted to DAMIT model 282. The average \textit{volume}-equivalent diameter from those 3 events is 102~km for DAMIT model 120, and 100~km for DAMIT model 282. These are slightly greater than the 98~km diameter determined by \cite{Sierks487}, but are consistent with the third axis being between 82 and 87~km  \textendash{} compared to the value of 75 $\pm{13}$ km found by \cite{Sierks487}.

    \begin{table}
	\centering
	\caption{Diameter of 21 Lutetia from occultation fits to shape models. The best events are asterisked. Columns D120 and D282 give the \textit{volume}-equivalent diameter (in km) when fitted to DAMIT models 120 and 282 respectively.}
	\label{Lutetia}
	\begin{tabular}{lccp{3.5cm}}
	\hline
	Date&D120&D282& Comment\\
	\hline
     2016 Sep 24 *&104&102& 2 well-spaced chords\\
     2017 Feb 10&110&100& 4 chords poorly distributed\\
     2019 Nov 25 *&102&100&8 chords evenly distributed\\
     2020 Jan 26 *&100&99&2 well-spaced chords\\
     2020 Mar 31&89-117&94-120&Poor fit. 4 well spaced chords with large timing uncertainties\\
     \hline
	\end{tabular}
\end{table}
    \begin{figure}
    	\includegraphics[width=\columnwidth]{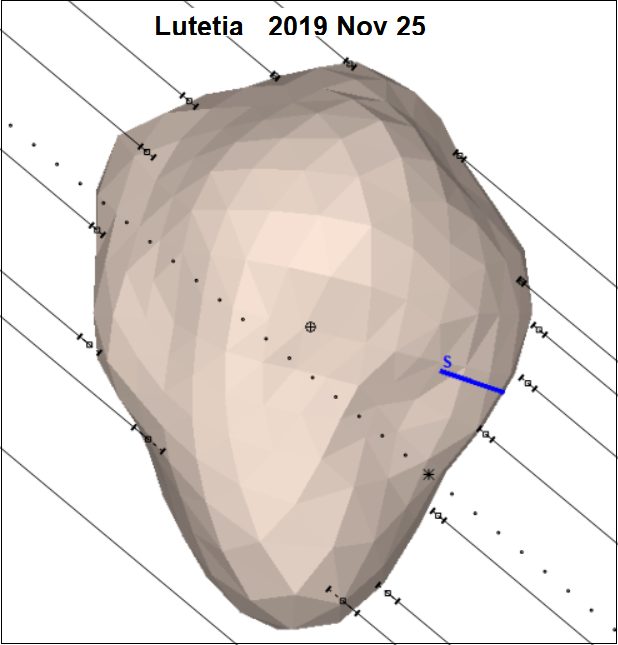}
        \caption{The fit of the chords from the 2019 Nov 25 occultation by (21) Lutetia against the shape model from \protect\cite{Lutetia282}. The relative motion of the star is from left to right.}
        \label{Lutitia}
    \end{figure}
    
    \item (433) Eros. The \textit{NEAR Shoemaker} probe orbited Eros in 2000. \cite{Yeomans2085} derived the volume of this very elongate asteroid as 2,503 $\pm{25}$~km$^3$, which gives a \textit{volume}-equivalent diameter of 16.8~km. Occultations by Eros typically have very short durations. Of the several observed occultations, events on 2011 Dec 13 and 2019 Mar 12 have the least time uncertainties. The results of those two occultations are shown in Fig. \ref{Eros}. While there are only two chords for each, their location is well-constrained by the shape of the asteroid. In both cases the derived \textit{volume}-equivalent diameter is 16.7~km, fully consistent with satellite-derived diameter. However the highly elongate nature of Eros prevents a meaningful determination of its diameter from a fitted ellipse.

    \begin{figure}
    	\includegraphics[width=\columnwidth]{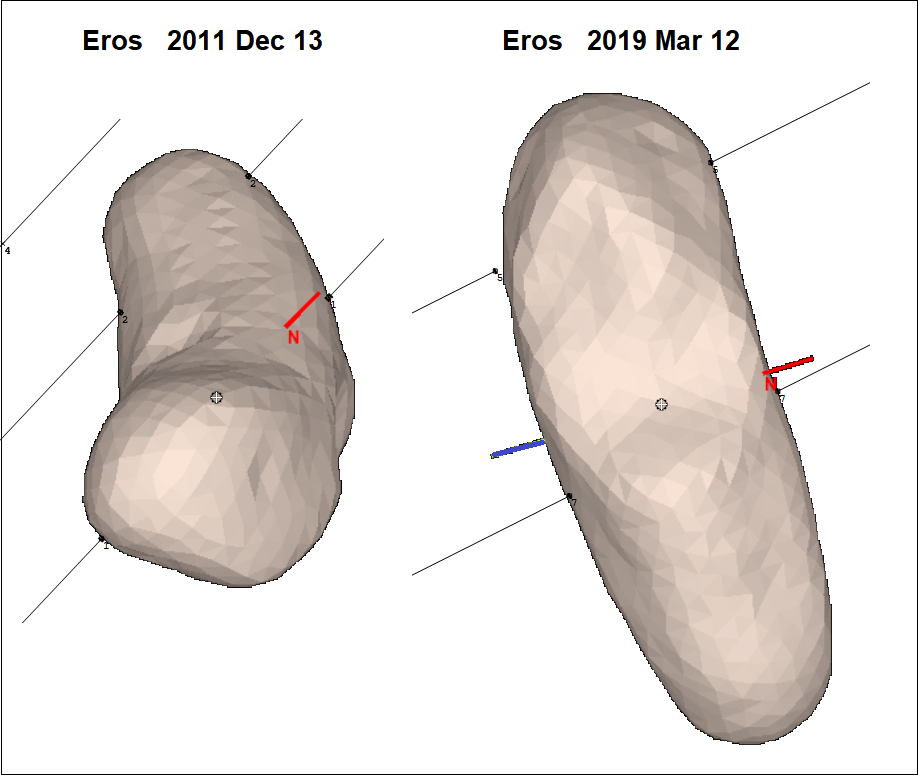}
        \caption{The fit of the chords from the 2011 Dec 13 and 2019 Mar 12 occultations by (433) Eros to the DAMIT shape model 3083 \protect(which is the 1,708 facets model from the \textit{NEAR} probe). The relative motion of the star is from left to right for both occultations. The occultation chord lengths for the 2011 Dec 13 event are 9.0 and 16.0~km; for the 2019 Mar 12 event they are 11.7 and 11.3~km.}
        \label{Eros}
    \end{figure}

    \item \label{486958} (486958) Arrokoth. 
    \textit{New Horizons} spacecraft flew past Arrokoth in January 2019.   \cite{2020Sci...367.3999S} report ellipsoidal axes of 36 x 20 x 10~km and a mean diameter of 18.3 km. \cite{Sterneaaw9771} report Arrokoth to be a contact binary with lobes of 22 x 20 x 7 km, and 14 x 14 x 10 km, giving a \textit{volume}-equivalent diameter of 17 km; they also provide a shape model\footnote{\textcolor{blue}{  https://3d-asteroids.space/asteroids/486958-Arrokoth}} with the axis of rotation being at ‘approximately' right ascension = 311\textsuperscript{o}, declination = –25\textsuperscript{o}, and a rotation period of 15.92 $\pm{0.02}$ h.

   An occultation by this asteroid on 2017 Jul 17 yielded 5 chords \citep{2020AJ....159..130B} -- which in Fig. \ref{arrokoth} are fitted against the shape model. This was 533 days before the flyby. From the rotation period there were 804 $\pm{1.01}$ rotations over this period; that is, the rotational orientation of the shape model derived from the \textit{New Horizons} spacecraft is indeterminate when referred to the date of this occultation. Using ecliptic coordinates for the axis of rotation ($\lambda$ = 306.8\textsuperscript{o}, $\beta$ = -6.6\textsuperscript{o}) the rotational phase $\gamma$ to give the fit in Fig. \ref{arrokoth} is 73.6\textsuperscript{o} at JD 2457951.0 (2017 July 16.5).

    \begin{figure}
    	\includegraphics[width=\columnwidth]{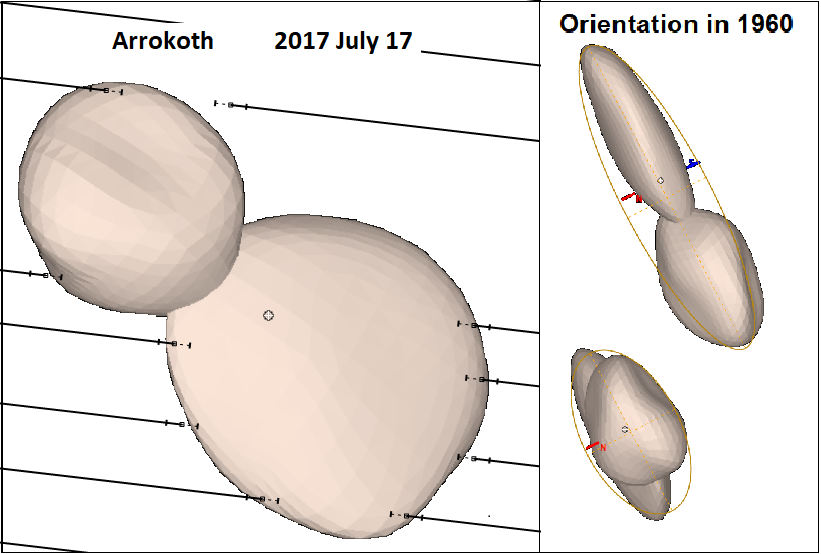}
            \caption{The left panel shows the chords from the 2017 July 17 occultation against the shape model of (486958) Arrokoth from \protect\cite{Sterneaaw9771}. The axis of rotation is approximately normal to the image. The right panel shows Arrokoth's orientation 4 hours apart in 1960, on the same month and day as the 2017 occultation, and with the axis of rotation shown. An indicative best-fit ellipse is overlaid. These two images are scaled at 60 per cent of the left image.}
        \label{arrokoth}  
   \end{figure}
 
   While most chords are well fitted to the profile, the uppermost chord is not. This chord could not be aligned with the shape model by variations of the polar axis coordinates of up to $\pm{16}\textsuperscript{o}$. However a fit consistent with all the other chords can be obtained by adding 0.10 seconds to the event times of that chord. This is not to say that there is a time base error with that chord. Rather it illustrates a situation where we would undertake a close review of the relevant observation to ensure the reported times are correct and reliable.
   
   The fit of the chords (as shown in the left panel) to the shape model gives a \textit{volume}-equivalent diameter of 16.7 $\pm{0.7}$ km, fully consistent with \cite{Sterneaaw9771}.

    As illustrated in Fig \ref{arrokoth}, Arrokoth is a highly irregular asteroid. The current orientation of the asteroid as shown in the left panel fails to disclose the relatively thin nature of the larger lobe -- such that any ellipse fitted to the profile will give a diameter much greater than its real diameter. Indeed a fitted ellipse has dimensions of about 34 x 18 km, giving a mean diameter of 25 km -- 50 per cent larger than its \textit{volume}-equivalent diameter.
    
    The right panel shows Arrokoth's orientation in 1960 on the same month and day as the 2017 occultation. The images are separated in time by 4 hours, and are overlaid with approximate best-fit ellipses. The ellipse of the upper image is 34 x 9 km, giving an equivalent diameter of 17 km - which happens to be the \textit{volume}-equivalent diameter of Arrokoth. The ellipse of the lower image is 18 x 10 km, giving a mean diameter of 13 km -- which is about 80 per cent of the \textit{volume}-equivalent diameter of Arrokoth. These three measures demonstrate that fitted ellipses do not provide a reliable measurement of the mean diameter of an irregular asteroid like Arrokoth.
\end{itemize}
    
\subsubsection{Mathilde - an asteroid without shape models}    

\begin{figure}
    	\includegraphics[width=\columnwidth]{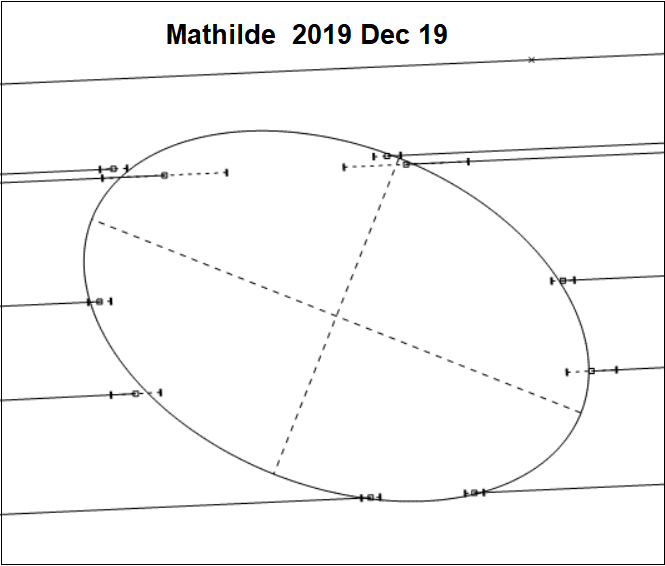}
        \caption{The fit of an ellipse to the chords of the 2019 Dec 19 occultation by (253) Mathilde. The relative motion of the star is from right to left.}
        \label{Mathilde}
    \end{figure}

(253) Mathilde. \cite{1999Icar..140....3V} gives the dimensions of Mathilde from the \textit{NEAR Shoemaker} flyby in 1997 as 66 x 48 x 44~km, with a \textit{volume}-equivalent diameter of 52.8 $\pm{2.6}$~km. Neither DAMIT nor ISAM have shape models for this asteroid. A 2019 Dec 19 occultation has five chords well distributed across the highly-elliptic profile, as shown in Fig \ref{Mathilde}. Using a reduced weight for the chord with large uncertainties, the ellipse fitted to the chords has axes of 68.1 $\pm{1.5}$ x 44.0 $\pm{1.4}$~km, for a mean diameter of 54.7 $\pm{2.1}$~km \textendash{} which is 2.6 per cent larger than the NEAR \textit{volume}-equivalent  diameter.\\

\textbf{Conclusion}
\newline While it might be thought obvious, the above comparisons establish that matching occultation chords from a single occultation event to a shape model gives a reliable \textit{volume}-equivalent diameter of the asteroid to within a few per cent at worst -- even if the asteroid's shape is highly irregular -- provided that:
\begin{itemize}
    \item the occultation results are reliable, particularly with respect to the event times and associated uncertainties;
    \item the observed chords are adequately distributed across the asteroid; and
    \item the shape model is reliable.
\end{itemize}
Also, the fit of an ellipse to the asteroid profile can give a reasonable result provided the asteroid is not too irregular.

\subsection{Diameters using different shape models}\label{DiasFromSM}
\begin{figure}
	\includegraphics[width=\columnwidth]{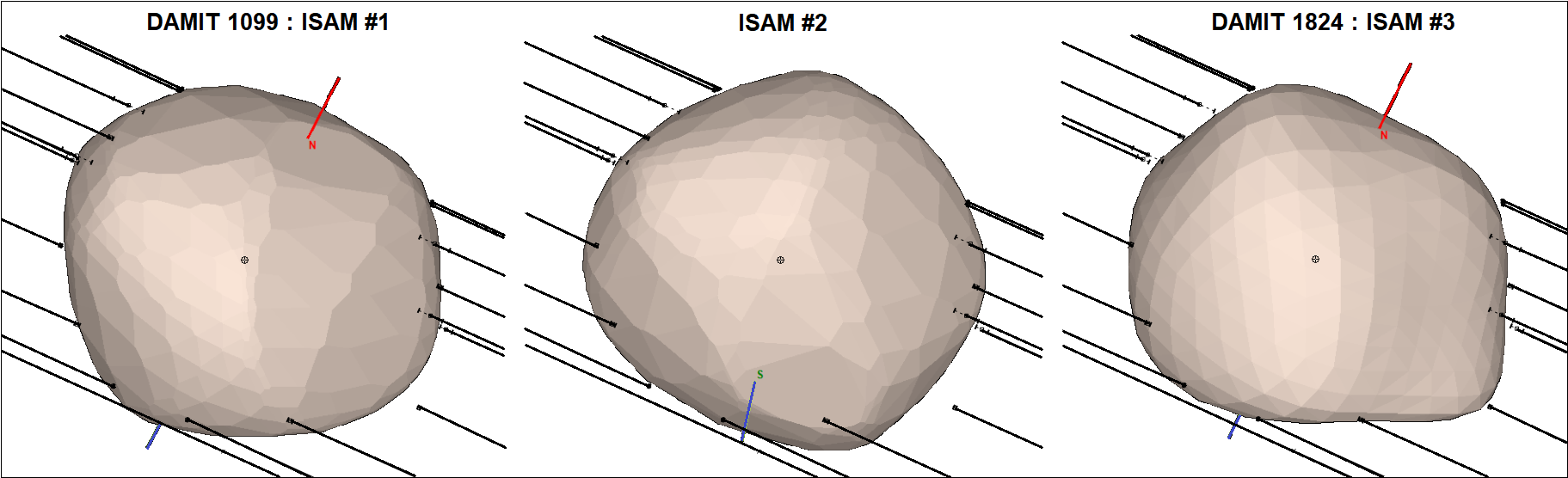}
    \caption{Three shape models fitted to the occultation by (144) Vibilia on 2011 Jan 25. The north pole is towards the top on the left and right images. The south pole is towards the bottom on the middle image. The lower 4 chords provide the primary discrimination between the models.} 
    \label{Vibilia}
\end{figure}

Fig.~\ref{Vibilia} shows the fit of three shape models to the occultation by (144) Vibilia on 2011 Jan 25. The shape models are from \protect\cite{Damit1099} [DAMIT 1099 = ISAM 1, and ISAM 2], and  \protect\cite{2017A&A...601A.114H} [DAMIT 1824 = ISAM 3]. Visual observations have been excluded to ensure all displayed chords are reliable. The orientation of the axis of rotation is shown for all three models. The differences between the three models can be attributed to:\begin{itemize}
    \item differences in the orientation of the axis of rotation; and
    \item the use of a concave solution for the right-hand model.
\end{itemize}
 The right-hand model provides a very good fit to all chords, which leads to a reliable determination of the diameter. The left and middle models do not have a good fit to the occultation chords, as seen by the bottom three chords. However, and in contrast to  Fig \ref{Comacina}, there are only a small number of chords that have a significant (but not very large) mismatch to the left and middle models, making it reasonable to determine the diameter of the asteroid using a best fit to the left and middle shape models. The diameters so determined are given in Table \ref{Vibilia_IR}, which have a very high level of consistency in this case. 

\begin{table}
	\centering
	\caption{Diameters of (144) Vibilia measured by the 2011 Jan 25 occultation fit to 3 shape models. }
	\label{Vibilia_IR}
	\begin{tabular}{rr}
	\hline
	Source& Dia (km) \\
	\hline
	 D1099 = ISAM 1&141 $\pm$1~km\\
	 ISAM 2&141 $\pm$3 km\\
	 D1824 = ISAM 3&142 $\pm$3~km\\
     \hline
	\end{tabular}
\end{table}

The conclusion we draw from this is that a reliable diameter can be determined whenever a majority of the observed chords can be reasonably matched to a shape model \textendash{} irrespective of whether differences are caused by inadequacies in the shape model, or observation errors in specific chords. The Quality setting in the data-set follows this principle. When there is a good correspondence between the observed chords and the shape model, the quality is set to \textit{Good}. When there are significant discrepancies but the chords can be reasonably fitted to the shape model, the quality is set to \textit{Poor}. When there are major discrepancies but the chords can be generally matched to the shape model, the quality is set to \textit{Diameter, but no fit}.

\subsection{Diameters from ellipse fits compared to shape model fits}
Shape models are only available for a portion of asteroids observed in an occultation. Where a shape model is not available, the only basis for determining an asteroid's diameter is by fitting an ellipse to the observed chords. The mean diameter of the asteroid is then the square-root of the product of the major and minor axes. With the profile of an asteroid generally varying in accordance with its rotational orientation, there is a real question of whether a well-constrained ellipse fitted to the occultation chords provides a reasonable measure of the diameter of an asteroid.

If an asteroid is essentially spherical (e.g. Ceres, in fig \ref{Ceres}) or a prolate spheroid (e.g. Vesta, in Fig \ref{Vesta}), the asteroid's profile and magnitude will have a low dependence on its rotational orientation, and a fitted ellipse may be expected to provide a good diameter measurement. In contrast, if an asteroid is highly irregular (e.g. Kleopatra in Fig \ref{figKleopatra}, or Eros in Fig \ref{Eros}), the asteroid's profile and magnitude will have a high dependence on its rotational orientation.

The irregularity of an asteroid might be inferred from the maximum amplitude of its magnitude variation. The greater the variation, the greater the irregularity. Light curve variations are conveniently available from the Asteroid Light Curve Database \footnote{\textcolor{blue}{ http://www.minorplanet.info/datazips/LCLIST\_PUB\_CURRENT.zip}\label{LCDB}} \citep{lcdb}, in the file LC\_SUM\_PUB.TXT which is held in LCLIST\_PUB\_CURRENT.zip. 

However light curve variations are only reliable if the asteroid has been observed at different polar orientations, requiring observations over a significant portion of its orbit. While this is probable for main-belt asteroids, it is presently unlikely for TNO's and Centaurs. The asteroid (486958) Arrokoth, illustrated in Fig. \ref{arrokoth}, is a case in point. \cite{Benecchi_2019} was unable to detect its highly irregular nature from photometry made over 9 days, because its pole vector was near the Earth's line of sight.

\begin{figure}
    \centering
	\includegraphics[width=.4\textwidth]{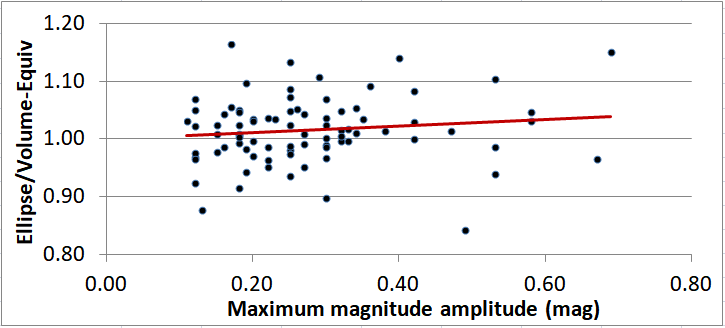}
    \caption{The ratio of elliptic fit diameter to the \textit{volume}-equivalent diameter from shape model fits, plotted against the maximum amplitude in light variation. The plot shows that the diameters derived from an ellipse fitted to chords is generally about 1.5 per cent larger than from a shape model fitted to the chords. The linear trend line is inadequate to suggest a dependency on asteroid irregularity, because it is poorly constrained.}
    \label{EllipseOverVolShape}
\end{figure}

\begin{figure}
  \centering
	\includegraphics[width=.4\textwidth]{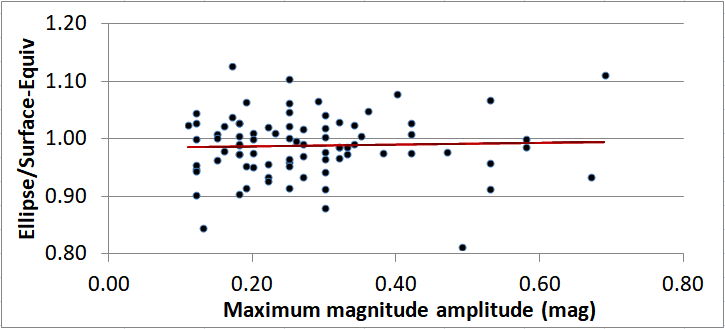}
    \caption{The ratio of elliptic fit diameter to the \textit{surface}-equivalent diameter from shape model fits, plotted against the maximum amplitude in light variation. The plot shows that the diameters derived from an ellipse fitted to chords is generally about 1.5 per cent smaller than from a shape model fitted to the chords. The poorly constrained linear trend line suggests no dependency on asteroid irregularity.}
    \label{EllipseOverSurfShape}
\end{figure}

To assess the dependency of ellipse-derived diameters against irregularity of an asteroid, we selected 83 occultations having a good fit to a shape model and a well-constrained fitted ellipse. The number of chords involved in these events varied between 3 and 132, with the majority involving fewer than 10. We excluded asteroids with a maximum magnitude amplitude greater than 0.8 (specifically, the extremely elongate asteroid Kleopatra).  Fig. \ref{EllipseOverVolShape} plots the ratio of the ellipse diameter to the \textit{volume}-equivalent diameter from the shape model -- against the maximum amplitude of light variation. Fig. \ref{EllipseOverSurfShape} is a similar plot using the \textit{surface}-equivalent diameter from the shape model. The lines of best fit in these plots indicate that the diameter derived from an ellipse does not have any significant dependence upon asteroid irregularity if the maximum magnitude amplitude is less than 0.7.

 The scatter in Figs \ref{EllipseOverVolShape} and \ref{EllipseOverSurfShape} is similar. The extent of that scatter is such that no material difference can be inferred from the apparent differences in the lines of best fit. As a result, an ellipse fitted to an occultation event cannot distinguish between a \textit{volume}-equivalent and \textit{surface}-equivalent diameter; rather the diameter can only be described as a \textit{mean} diameter.

Combining this with the results from Subsection \ref{IRverify}, we conclude that if the maximum amplitude in light variation is less than 0.7 (or possibly somewhat larger), a well-constrained ellipse fitted to a single occultation event will provide a \textit{mean} diameter with a 1-sigma deviation of about 8 per cent, with that value rarely being in error by more than 10 per cent. Of course, a statistical combination of ellipse diameters from multiple events will improve the accuracy of the diameter measurement. 

\begin{figure}
\twocolumn
\centering
\includegraphics[width=.37\textwidth]{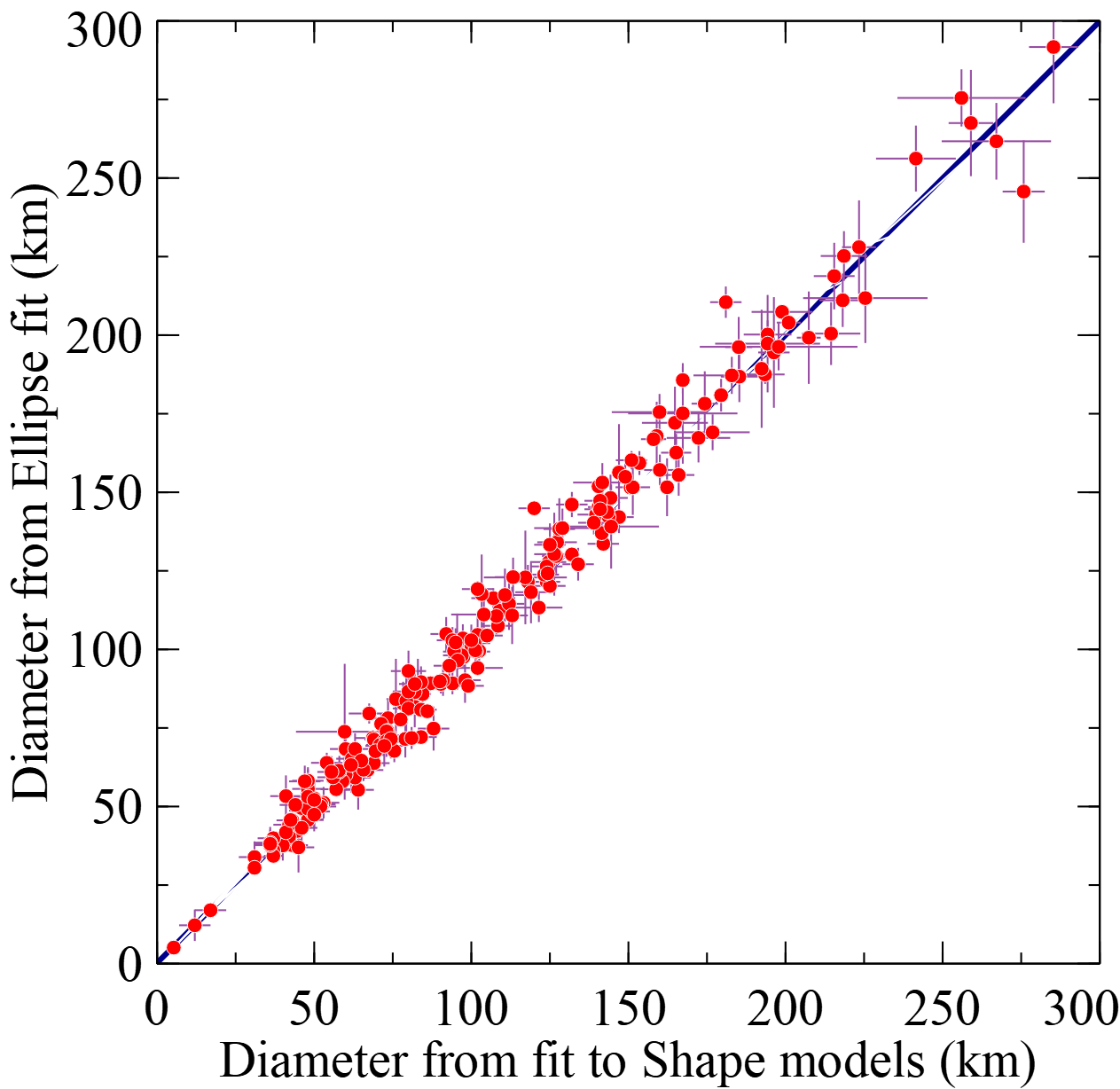}
    \caption{The diameter derived from a fitted ellipse against the \textit{volume}-equivalent diameter from a shape model, showing that a fitted ellipse is a good approximation to the \textit{volume}-equivalent diameter.}
    \label{EllipseModel}
\end{figure}

The validity of using ellipse fits to determine the \textit{volume}-equivalent diameter can also be assessed by plotting the ellipse fit diameter against the \textit{volume}-equivalent diameter from a shape model, as in Fig \ref{EllipseModel}. The plot is based on the 215 asteroids which had 
\begin{itemize}
 \item diameters determined from one or more shape models, and
 \item diameters from ellipse fits where:
  \begin{itemize}
    \item[\textbf{a.}] the quality of the fit to the chords was one of ‘Limits on size but no shape', ‘Reliable size. Can fit to shape models', or ‘Resolution better than shape models'; and
   \item[\textbf{b.}] there were 2 or more chords - a criterion largely redundant over the former.
  \end{itemize}
\end{itemize}

Both the shape model diameters and the ellipse diameters are plotted on the basis of their average diameters from multiple occultations, and corresponding standard deviations.

Fig \ref{EllipseModel} demonstrates that an ellipse fit can result in a reliable diameter determination for those asteroids not having a shape model. This is also illustrated by the analysis of two occultations of Ceres \citep{Gomes-Junior} where an equatorial diameter of 972 $\pm$6~km and an oblateness of 0.08 $\pm$0.03 was derived, giving a \textit{volume}-equivalent diameter of 945 $\pm$10km. This may be compared to the values of 937 $\pm$3~km and 937 $\pm$9~km found by fitting occultation chords to a shape model (Subsection \ref{CeresRef}), and the \textit{volume}-equivalent diameter from the \textit{Dawn} spacecraft of 937 $\pm$3~km. 

Shape model results are inherently more reliable than a diameter determination using a fitted ellipse. Accordingly we have limited all subsequent analysis to diameters determined by fitting to a shape model.

\subsection{Comparison with Earth-orbiting satellites}
\begin{figure}
  \centering
  \includegraphics[width=0.37\textwidth]{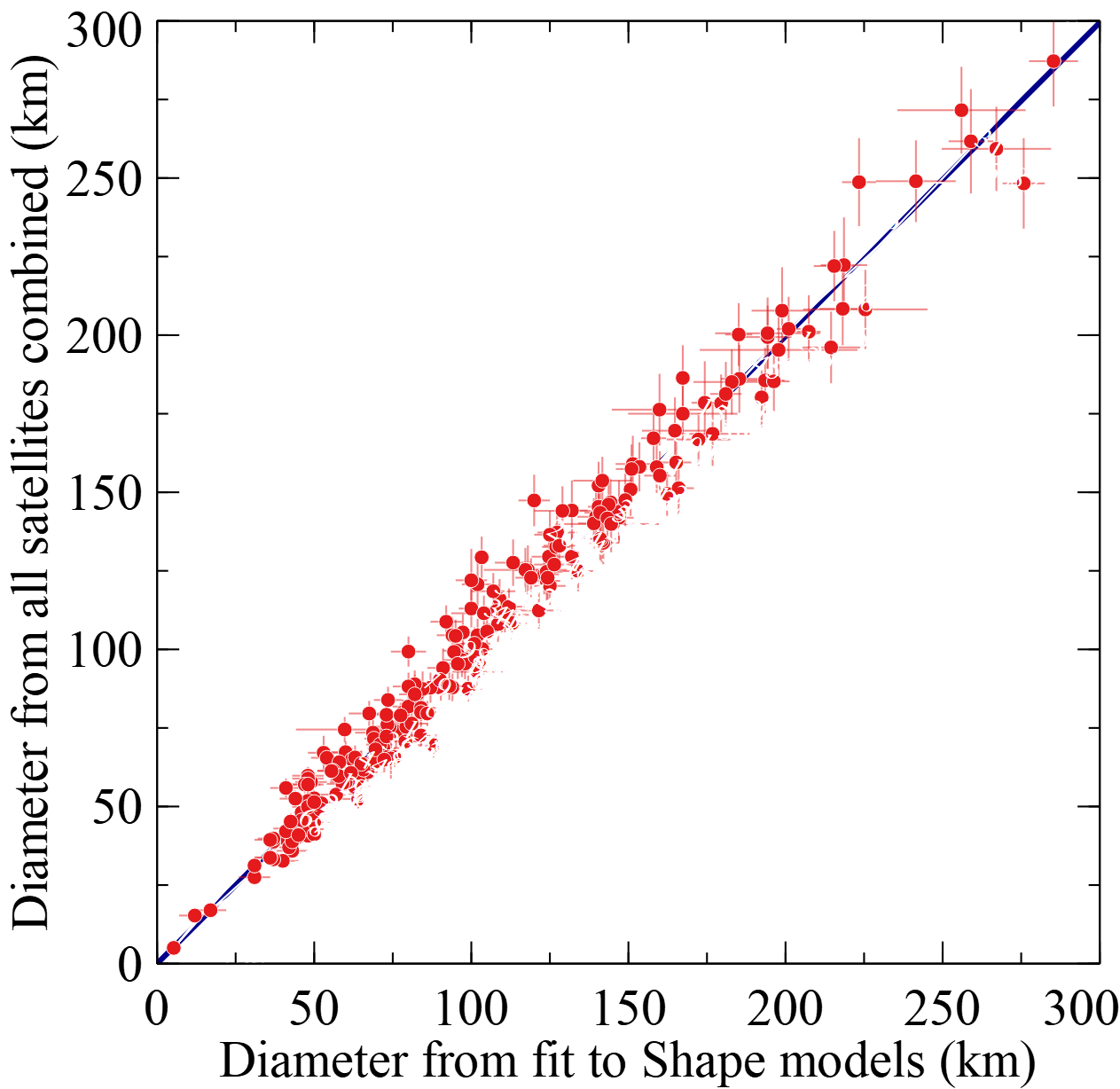}
  \caption{Diameters from weighted mean of all satellite measures vs. averaged \textit{volume}-equivalent diameters from shape model fits.}
  \label{fitAll}
\end{figure}

\begin{figure}
  \centering
  \includegraphics[width=0.37\textwidth]{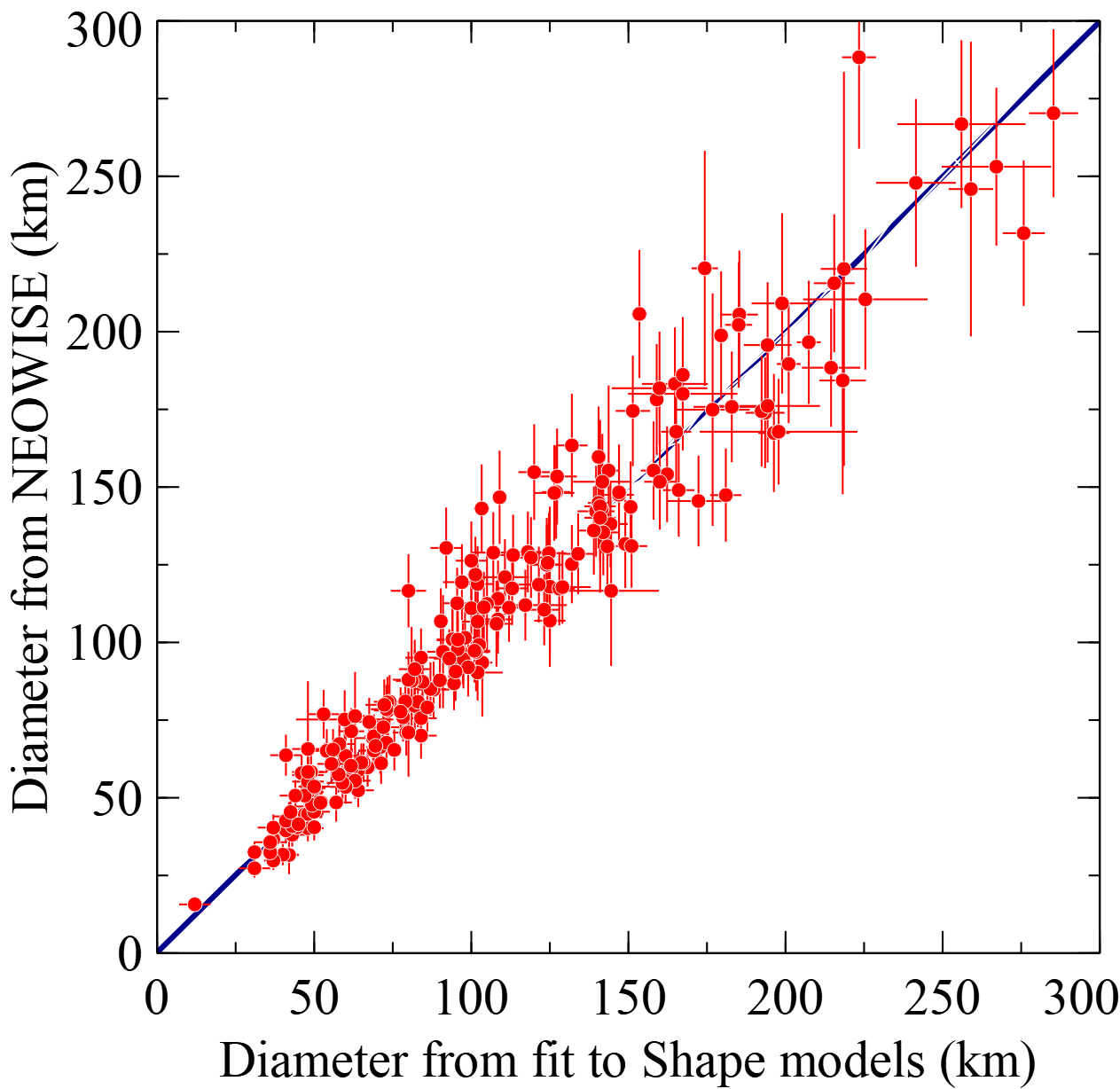}
  \caption{Diameters from weighted mean of individual \textit{NEOWISE} diameters vs. averaged \textit{volume}-equivalent diameters from shape model fits.}
  \label{fitNeo}
\end{figure}

\begin{figure}
  \centering
  \includegraphics[width=0.37\textwidth]{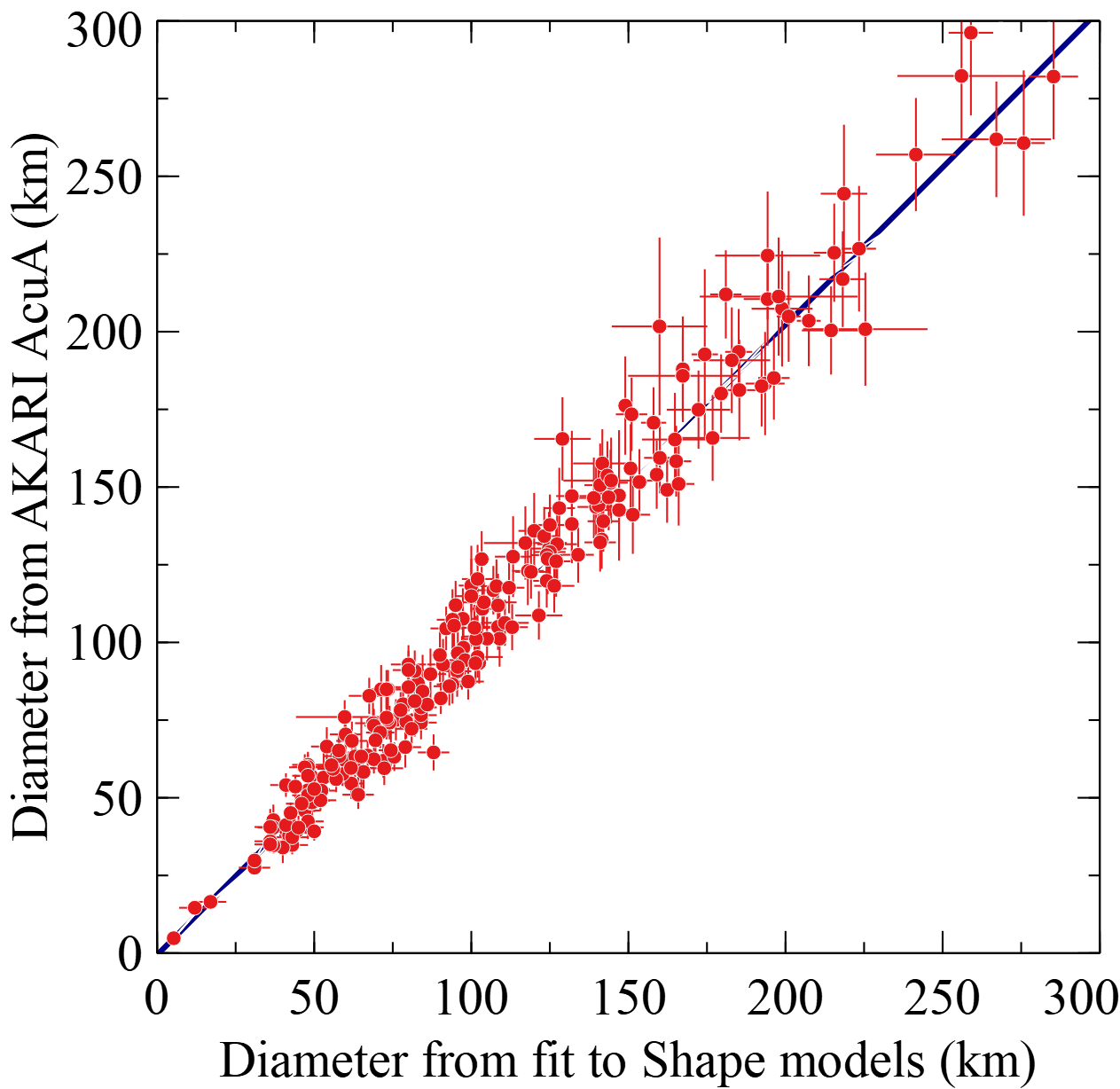}
  \caption{Diameters from weighted mean of individual \textit{AKARI AcuA} diameters vs. averaged \textit{volume}-equivalent diameters from shape model fits.}
  \label{fitAcuA}
\end{figure}

\begin{figure} 
  \centering
  \includegraphics[width=0.37\textwidth]{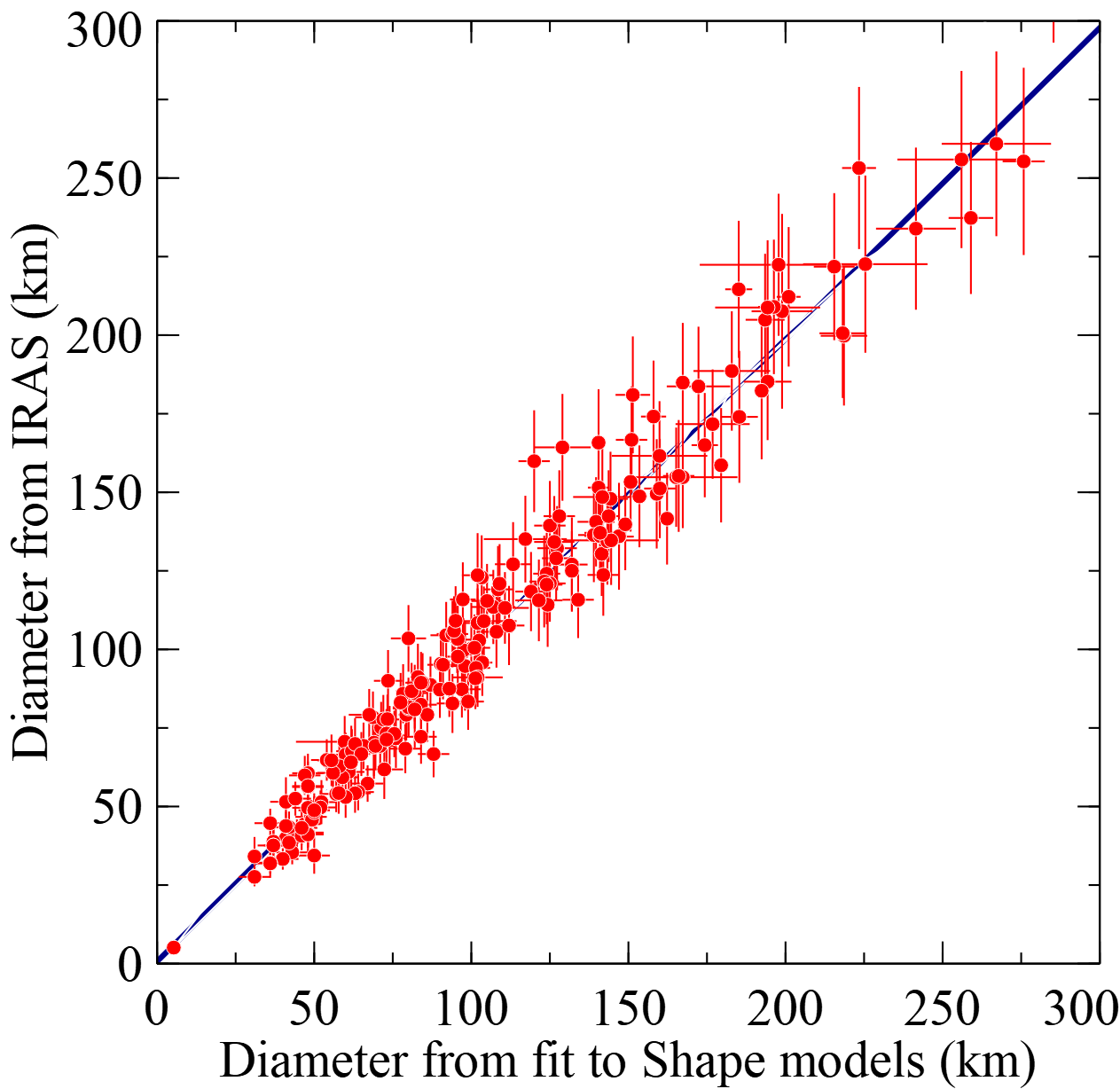}
  \caption{Diameters from weighted mean of individual \textit{IRAS} diameters vs. averaged \textit{volume}-equivalent diameters from shape model fits.}
  \label{fitIRAS}
\end{figure}

Three satellite missions have measured asteroid diameters -- \textit{NEOWISE}, \text{AKARI AcuA}, and \textit{IRAS}. The source of the diameters measured by these satellites is set out in Section 4. Figs \ref{fitAll}, \ref{fitNeo}, \ref{fitAcuA} and \ref{fitIRAS} plot respectively the combined measurements from all satellites and the combined measurements from each satellite, against the averaged \textit{volume}-equivalent diameters from occultations.

For this purpose we have used 219 asteroids having one or more good or poor fits to a DAMIT shape model, and used the mean of those diameters. We have also included a nominal standard deviation on the basis of the variations in those fitted diameters, using a fixed value of 5~km for the 19 asteroids which had only one fit to a shape model. 

For the plot involving all satellites we used the weighted mean of the individual diameter measurements from all three satellites, together with the corresponding standard deviation.

For the plots involving each individual satellite, we used the weighted mean of the individual diameter measurements by that satellite, together with the corresponding standard deviation.

The uncertainties listed in the table of diameters in \textit{NEOWISE} and \textit{IRAS} required adjustment. As explained in \cite{Mainzer_2011} and \cite{Neowise}, Bundle Description, Caveats -- a systematic error of 10 per cent needs to be added to the diameter uncertainties from the \textit{NEOWISE} observations. The 10 per cent uncertainty was generally applied to \textit{IRAS} diameters because the diameter determination technique was very similar to that used by \textit{NEOWISE}, with it being 15 per cent when there were fewer than 4 detections [to account for observational selection effects]. (Masiero, J., private communication). In both cases this additional uncertainty was added in quadrature to the catalogue uncertainty after each of the individual measurements and uncertainties were combined. These plots indicate that the diameter determinations are consistent with the occultation diameters to within the uncertainties specified for the relevant satellites. \label{IRASuncert}

Curiously, each of  Figs \ref{fitAll}, \ref{fitNeo}, \ref{fitAcuA} and \ref{fitIRAS} show a small increase above the line of best fit at around 100km, tapering off at about 150 km. The explanation for this is not apparent. Apart from this, the line of best fit indicates a relationship between the various satellite diameters and the occultation diameters that is essentially 1:1. While this might be expected, it does provide independent verification using a large sample of asteroids, of the validity (within the measurement uncertainties) of asteroid diameters measured by these satellites.

The scatter shown in Fig  \ref{fitAll} is clearly \textit{much} less than that of the individual satellite measurements -- demonstrating that the most reliable determination of diameter from these Earth-orbiting satellites is obtained by combining the available results from each.

\begin{figure}
  \centering
  \includegraphics[width=0.37\textwidth]{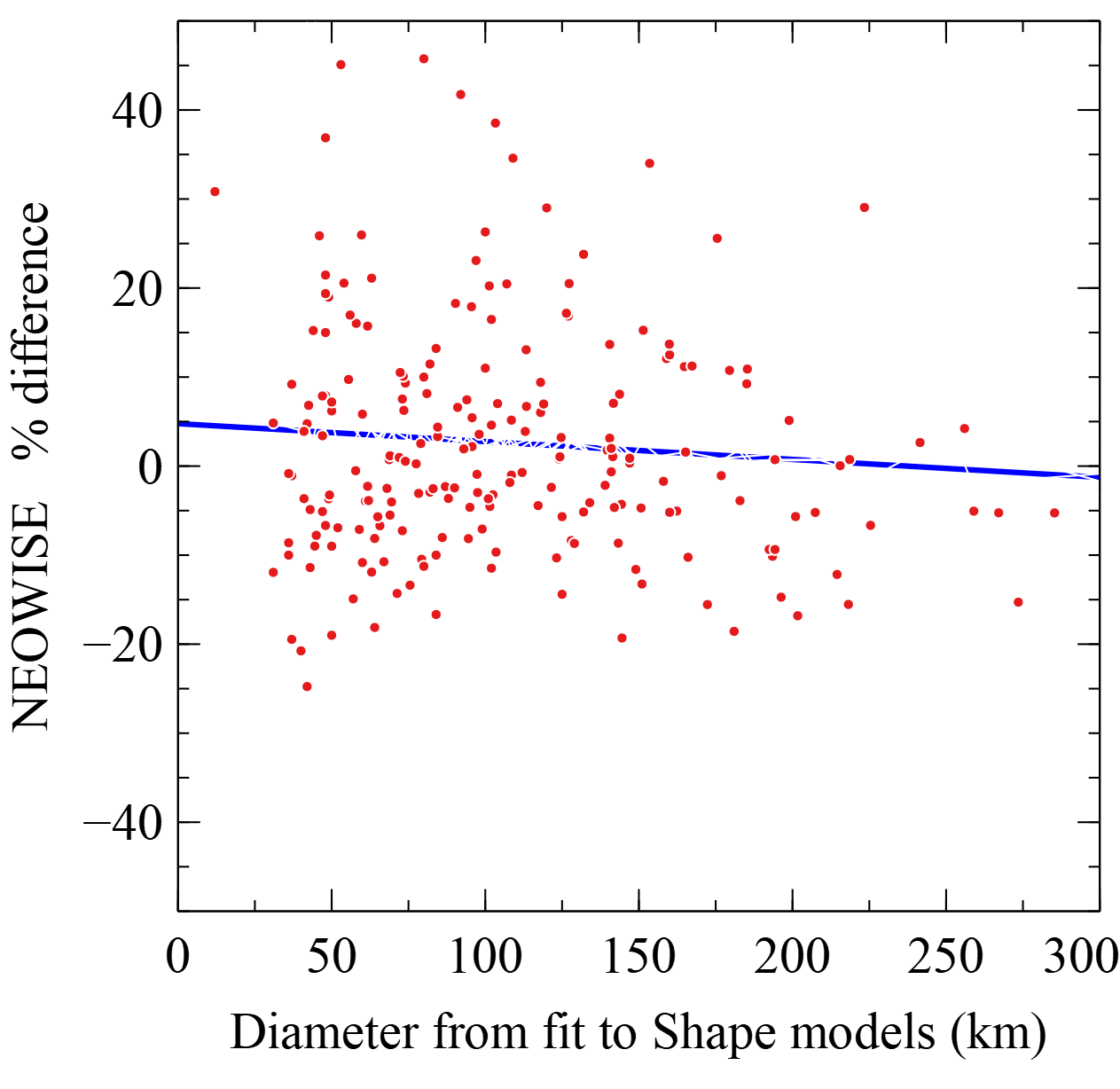}
  \caption{Per cent difference of individual \textit{NEOWISE} diameters against averaged \textit{volume}-equivalent diameters from shape model fits -- together with a line of best fit.}
  \label{PctA}
\end{figure}

\begin{figure}
  \centering
  \includegraphics[width=0.37\textwidth]{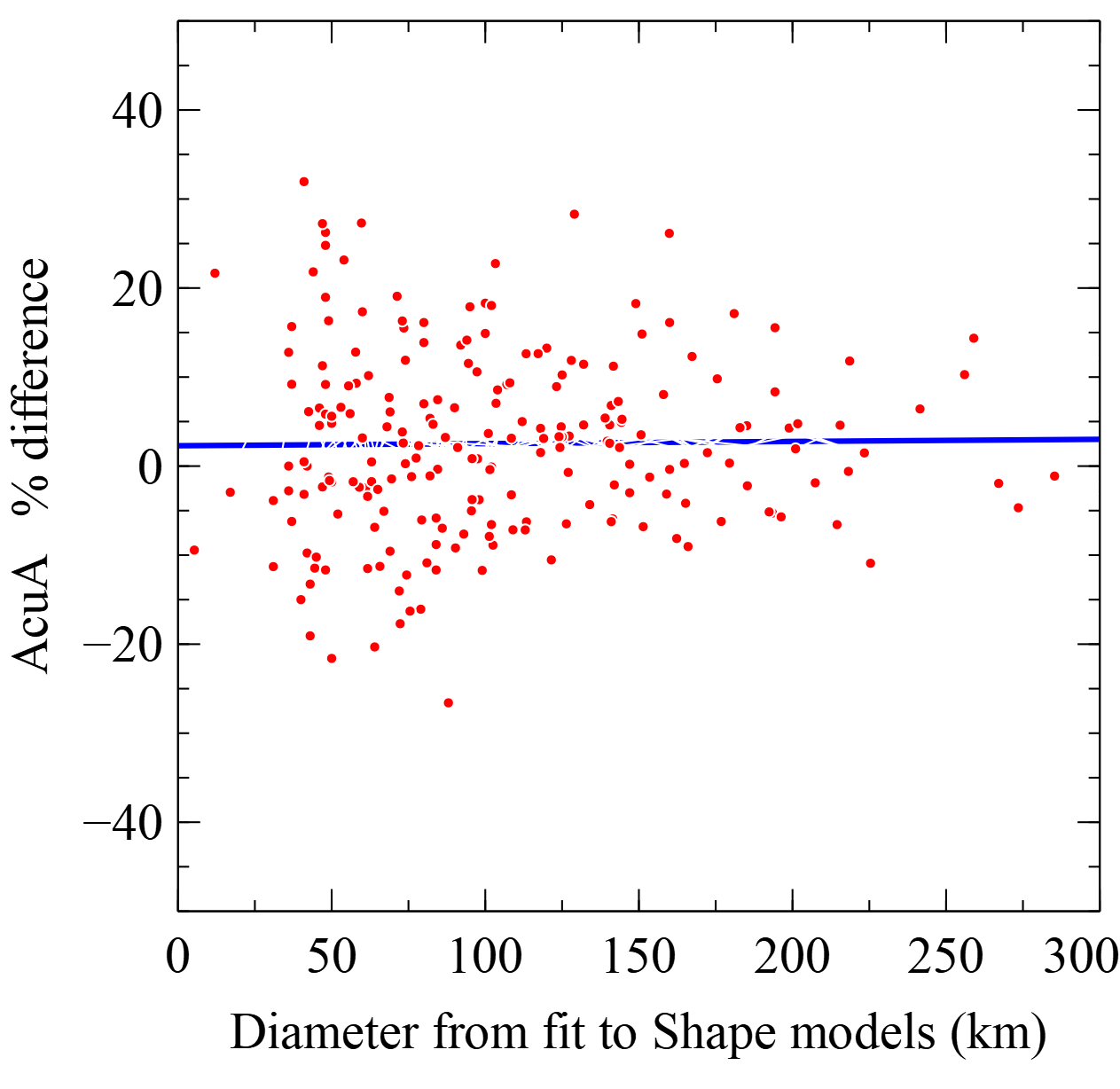}
  \caption{Per cent difference of individual \textit{AKARI AcuA} diameters against averaged \textit{volume}-equivalent diameters from shape model fits -- together with a line of best fit.}
  \label{PctB}
\end{figure}

\begin{figure}
  \centering
  \includegraphics[width=0.37\textwidth]{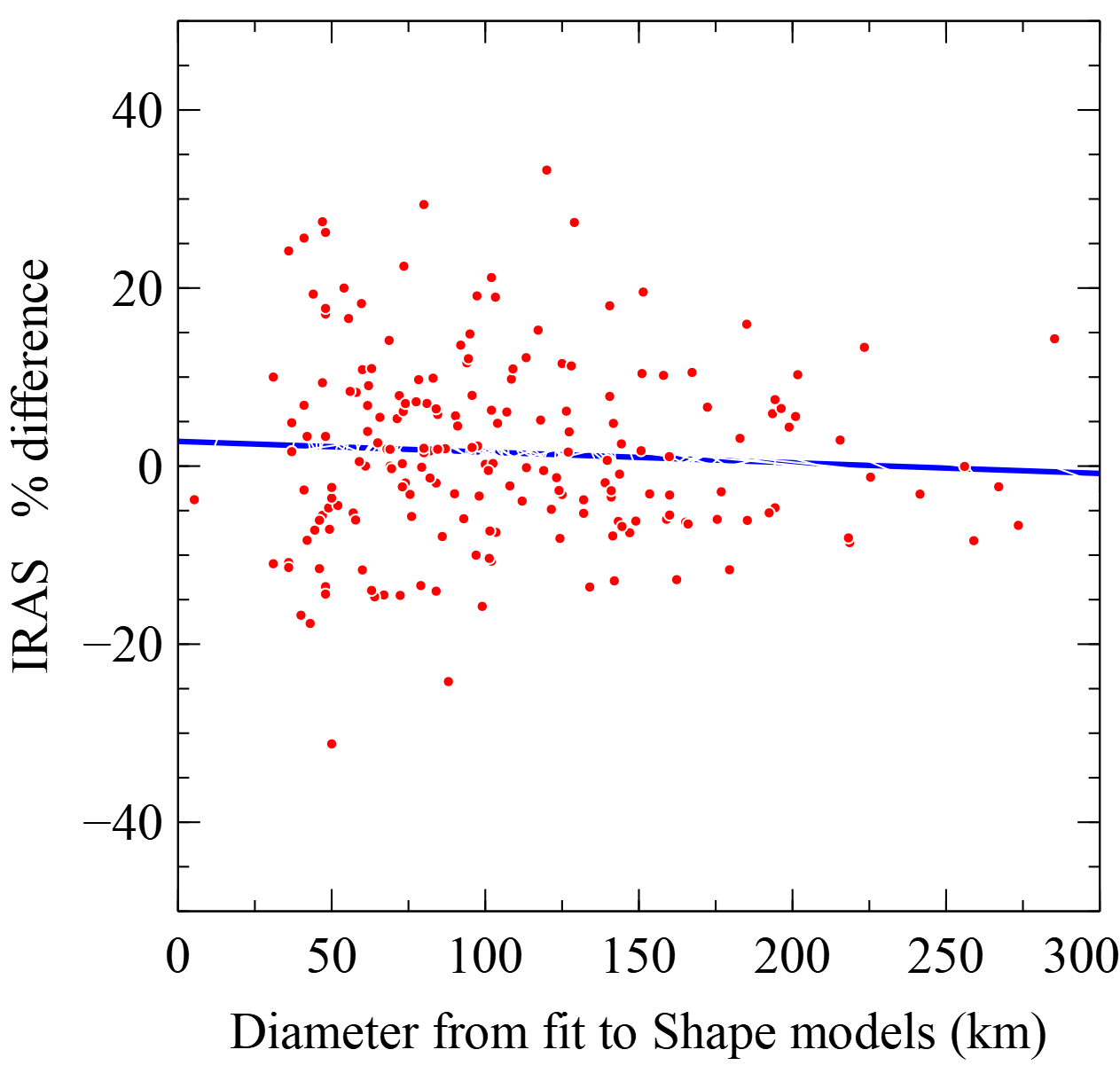}
  \caption{Per cent difference of individual \textit{IRAS} diameters against averaged \textit{volume}-equivalent diameters from shape model fits -- together with a line of best fit.}
  \label{PctC}
\end{figure}

Figs \ref{fitAll}, \ref{fitNeo}, \ref{fitAcuA} and \ref{fitIRAS} show that the absolute difference between measurements varies according to diameter. To show the differences in a manner that is independent of asteroid diameter, we plot in Figs. \ref{PctA}, \ref{PctB} and \ref{PctC} the percentage difference between the diameters measured by the various satellites and the occultation diameter. For information only, the plots include a line of best fit; given the scatter in these plots, any apparent trend in these 3 lines is not significant. These figures suggest that the percentage differences are larger for small asteroids. This is consistent with typical observing techniques with occultations, where the event is monitored with a fixed-interval video camera; a 0.1 second uncertainty in duration gives a much larger uncertainty in diameter for a 2 second occultation (typical of smaller asteroids), than for a 20 second occultation (typical of larger asteroids).  The consistency of this amongst the three satellites is consistent with this effect.

Points at the upper and lower extremes of these plots relate to asteroids having large percentage differences from the occultation diameters. Tables \ref{TablePlus} and \ref{TableMinus} identify the five asteroids having the greatest percentage difference -- both larger and smaller -- for each satellite.
\begin{table}
	\centering
	\caption{List of the 5 asteroids in Figs \ref{PctA}, \ref{PctB} and \ref{PctC} having the largest positive per cent difference in diameter compared to occultation diameters, in decreasing order of difference.}  
	\label{TablePlus}
	\begin{tabular}[3]{|r|r|r|}
	\hline
	 \textit{NEOWISE} & \textit{AcuA} & \textit{IRAS}\\
     \hline
	 679 &679&209\\
	 791 &1437&791\\
	 199 &44&816\\
	  98 &816 &1437\\
	 328 &218&218\\
    \hline
	\end{tabular}
\end{table}

\begin{table}
	\centering
	\caption{List of the 5 asteroids in Fig \ref{PctA}, \ref{PctB} and \ref{PctC} having the largest negative per cent difference in diameter compared to occultation diameters, in decreasing order of difference.}  
	\label{TableMinus}
	\begin{tabular}[3]{|r|r|r|}
	\hline
	 \textit{NEOWISE} & \textit{AcuA} & \textit{IRAS}\\
     \hline
	 302 &55&543\\
	 430 &695&55\\
	 732 &371&158\\
	 690 &158 &430\\
	 695 &849&528\\
    \hline
	\end{tabular}
\end{table}

Tables \ref{TablePlus} and \ref{TableMinus} show little commonality of these asteroids as between the three satellites. Such differences cannot be caused by any issues with the occultation diameters. Rather they are associated with the relevant satellite measurements.

\subsection{Occultation diameters -- a list of the best}
Table \ref{Diameters} provides a comparative list of the best diameters derived from asteroidal occultations, limited to those which: 
\begin{itemize}
    \item have been fitted to shape models
    \item the quality of the occultation event having being assessed as either reliable, or having greater detail than the shape model
    \item are restricted to events having at least 3 chords (which is largely superfluous over the quality requirement), and
    \item there being at least two such events for the asteroid.
\end{itemize}

The uncertainty of the shape model occultation diameters listed in Table \ref{Diameters} is based on the difference in diameters from different events, rather than the uncertainty associated with individual events. The uncertainty can be attributed to difficulties in matching occultation chords to shape models, occultation chords fitting two or more shape models with potentially different diameter determinations (subsection \ref{DiasFromSM}), occultation timing uncertainties, and the accuracy of shape models. Nevertheless, as shown in Table \ref{Diameters}, the uncertainty of occultation diameters is generally much smaller than satellite uncertainties.

\section{The Future}\label{Future}
The observation of asteroidal occultations has continuously improved since the first successful observation in 1961 Oct 2 -- an occultation by Pallas observed photoelectrically and visually from India \citep{Pallas}. As previously noted in several parts of this paper, the content of the data-set is continuously being improved, as well as new events being added. The following is a list of expected improvements in both the data-set and observations that might be expected over the next few years:
\begin{itemize}
  \item Ongoing improvements to the data-set as needs and circumstances evolve. For example, there is a need to deal with the effects of asteroid rotation between the event times of observers who are widely spaced, such as intercontinental.
  \item\textit{Gaia} Data Release 2 considered asteroid orbits as part of the validation process \citep{Gaia2}, with some of the orbits having a quality equivalent to the better ground-based observations. Data Release 3 is expected to include orbits of asteroids with much greater accuracy. We hope there will be an improvement in orbital elements similar to the improvements in star positions that \textit{Gaia} has provided. If that is the case, the uncertainty in the predicted location of an occultation path will be reduced to just a few km (the precision being limited by the accuracy of the geocentric position of the \textit{Gaia} spacecraft when it measures an asteroid's position), enabling reliable predictions. This will provide certainty for fixed observers. It will also enable groups of observers to confidently travel to the location of a predicted path and spread themselves across the path -- thereby measuring the profile and size of many more asteroids than is currently the case. And it will make it practical to observe occultations by smaller asteroids -- down perhaps to 1~km in diameter, assuming the occultation duration is not too short. [Already 6 occultations of the 5~km asteroid (3200) Phaethon have been observed.]
 \item One observing arrangement is for a single observer to deploy multiple portable telescopes alongside roads. Such telescopes are fixed, pre-pointed to where the star will be at the time of the occultation, and have a video camera recording the image as the star field drifts across. Improvements continue to be made to increase their aperture and ability to be quickly deployable, whilst being able to fit many such telescopes in a car.
  \item Electronic detectors continue to be improved. Given the number of active observers involved in asteroidal occultations, it is inevitable that new detectors will be tried by someone, with results/recommendations being shared.
  \item The most critical item in an observation setup is that of accurate time. Without accurate time, the observations of different observers cannot be reliably combined. The most commonly used time source involves a GPS receiver which places a time stamp on each field of an analogue video stream. However there is ongoing investigation of other ways to link accurate time (to within 10msec or less). For example, linking time to a Digital Video recording made using modern CMOS cameras; using properly calibrated Network Time Protocol (NTP) \citep{Pavlov}; calibrating USB Bus Latency when the PC is used to time-stamp the exposure; and ensuring accurate timing methods are available for mobile observers setting up multiple stations.
\end{itemize}

\section{Conclusions}
We have published and archived at NASA's Planetary Data System\textsuperscript{\ref{PDS}} a data-set of all asteroidal occultation observations. We outline the data-set, and how the observations were reduced. To illustrate what can be derived from this data-set we:
\begin{itemize}
\item show how high precision astrometry, referred to the centre of mass or figure, is derived from an occultation. 
\item explore the issues associated with combining occultation results with shape models or fitted ellipses, to obtain a reliable measurement of an asteroid's diameter.
\item compare diameters from occultations with diameters measured by visiting spacecraft, and confirm that a single well-observed occultation event can accurately measure the \textit{volume}-equivalent diameter of an asteroid.
\item illustrate the accuracy obtained when a binary asteroid or an asteroidal satellite occults a star, and determine the mean diameter of (87) 1 Romulus to be 25 $\pm$1 km.
\item provide a list of 76 double stars discovered in occultations, with separations typically in the tens of mas or less.
\item list the diameter, separation and position angle of 30 asteroidal satellites and possible satellites.
\item compare the occultation diameters with the diameters from three Earth-orbiting satellites; demonstrate general consistency between occultation diameters and satellite measurements; show that the uncertainty in occultation diameters is generally less than the uncertainty in satellite diameters; and show that the best satellite-determined diameters are obtained by combining the measurements from all three satellites.
\item list 104 asteroids having the most reliable occultation diameters, together with the diameters from each of the three Earth-orbiting satellites for comparison. 
\end{itemize}

Overall, we believe this data-set provides a valuable source of information relating to high-precision astrometry, and diameters, of asteroids.

\section*{Acknowledgements}
We thank the referee (Josef Durech) for his very helpful comments for improving the paper.

We thank Przemyslaw Bartczak for converting the \textit{Dawn} shape model data for Ceres and Vesta into the format used in DAMIT and ISAM shape model websites, and Breno Loureiro Giacchini for some drafting advice.

As previously noted, the data-set contains over 15,000 observations by more than 3,300 individuals from around the world, involving more than 4,400 events over a period greater than 40 years. The great majority of observers have made these observations at their own expense, including occasions when they have traveled significant distances. Most of those observers are affiliated with one or more of:
\begin{itemize}
    \item European Asteroidal Occultation Network (EAON)
    \item International Occultation Timing Association (IOTA)
    \item International Occultation Timing Association -- European Section (IOTA--ES)
    \item Japanese Occultation Information Network (JOIN)
    \item Trans Tasman Occultation Alliance (TTOA) \end{itemize}
and we thank them for their contributions.

\begingroup
    \courier
    \small
    \begin{center}
    \onecolumn
    \begin{longtable}[h]{rlrlrrlrlrl}

         \caption{Comparison of occultation diameters with diameters from \textit{NEOWISE}, \textit{AKARI AcuA}, and \textit{IRAS}\label{comparison}}\\

         \multicolumn{11}{l}{104 diameters (km) derived from the fit of occultation observations to shape models, together with the}\\
         \multicolumn{11}{l}{diameters determined by the \textit{NEOWISE}, \textit{AKARI AcuA}, and \textit{IRAS} satellites. The uncertainties for NEOWISE}\\
         \multicolumn{11}{l}{and IRAS include the 10 per cent increase referred to at \ref{IRASuncert}}\\
         \multicolumn{11}{l}{}\\
         \multicolumn{11}{l}{}\\
         \multicolumn{11}{l}{The asteroids listed have at least 2 occultations events, each event having at least 3 chords and an event}\\
         \multicolumn{11}{l}{quality of either ‘Reliable size. Can fit to shape models', or ‘Resolution better than shape models'.}\\
         \multicolumn{11}{l}{The column ‘Shape Mod' gives the mean and standard deviation of the diameters derived from occultation events,}\\
          \multicolumn{11}{l}{with the number of events used being listed in the column ‘$\sharp$'.}\\
         \multicolumn{11}{l}{}\\

         \hline
         Number & Name        &Shape Mod.&Uncert  &$\sharp$ & \textit{NEOWISE}& Uncert & \textit{AcuA} &Uncert& \textit{IRAS} &Uncert     \\
         \hline
         \endfirsthead
         
         \hline \multicolumn{4}{l}{\textit{Continued on next page}} \\
         \endfoot
        
        \multicolumn{3}{l}
        {\textit{Continuation of}\ \tablename\ \thetable}\\
        \hline
        Number & Name        &Shape Mod.&Uncert  & $\sharp$ & \textit{NEOWISE}& Uncert & \textit{AcuA} &Uncert& \textit{IRAS} &Uncert     \\
       \hline 
        \endhead
         
        \hline
        \endlastfoot

     2 &Pallas       &   527.7 & $\pm$ 18.6& 10&   544.0 & $\pm$ 67.0&   545.8 & $\pm$ 38.6&   498.1 & $\pm$ 53.4 \\
     3 &Juno         &   241.5 & $\pm$ 12.7&  4&   247.9 & $\pm$ 27.0&   257.0 & $\pm$ 18.2&   233.9 & $\pm$ 25.8 \\
     5 &Astraea      &   113.0 & $\pm$  1.0&  2&   107.4 & $\pm$ 11.0&   105.0 & $\pm$  7.4&   119.1 & $\pm$ 13.8 \\
     6 &Hebe         &   194.3 & $\pm$  7.6&  3&   195.7 & $\pm$ 20.2&   210.5 & $\pm$ 15.0&   185.2 & $\pm$ 18.6 \\
     7 &Iris         &   213.0 & $\pm$  3.7&  3&   220.2 & $\pm$ 63.4&   244.4 & $\pm$ 22.2&   199.8 & $\pm$ 22.2 \\
     8 &Flora        &   147.0 & $\pm$  4.7&  4&   147.5 & $\pm$ 15.0&   142.6 & $\pm$ 10.2&   135.9 & $\pm$ 13.8 \\
     9 &Metis        &   164.8 & $\pm$ 10.5& 14&   183.2 & $\pm$ 18.2&   165.3 & $\pm$ 15.0&    .... &            \\
    10 &Hygiea       &   423.0 & $\pm$  2.0&  2&   453.2 & $\pm$ 49.4&   411.4 & $\pm$ 37.0&   407.1 & $\pm$ 41.4 \\
    11 &Parthenope   &   150.7 & $\pm$  0.9&  3&   143.6 & $\pm$ 14.6&   156.0 & $\pm$ 14.2&   153.3 & $\pm$ 15.8 \\
    13 &Egeria       &   198.9 & $\pm$  9.7& 10&   209.1 & $\pm$ 29.0&   207.4 & $\pm$ 18.6&   207.6 & $\pm$ 31.0 \\
    16 &Psyche       &   223.0 & $\pm$  6.0&  5&   288.3 & $\pm$ 29.4&   226.7 & $\pm$ 20.2&   253.2 & $\pm$ 25.8 \\
    17 &Thetis       &    73.5 & $\pm$  4.5&  2&    78.1 & $\pm$  8.2&    84.9 & $\pm$  6.2&    90.0 & $\pm$  9.8 \\
    18 &Melpomene    &   138.4 & $\pm$  6.3&  5&   142.2 & $\pm$ 14.6&   143.6 & $\pm$ 13.0&   140.6 & $\pm$ 14.2 \\
    19 &Fortuna      &   206.3 & $\pm$  3.6&  6&   196.6 & $\pm$ 19.8&   203.5 & $\pm$ 14.6&    .... &            \\
    21 &Lutetia      &   103.7 & $\pm$  3.7&  6&    93.5 & $\pm$ 17.4&   110.8 & $\pm$  7.8&    95.8 & $\pm$ 10.6 \\
    22 &Kalliope     &   151.4 & $\pm$  5.5& 10&   174.5 & $\pm$ 17.8&   141.1 & $\pm$ 12.6&   181.0 & $\pm$ 18.6 \\
    25 &Phocaea      &    71.3 & $\pm$  0.9&  3&    61.1 & $\pm$  6.6&    84.9 & $\pm$  7.8&    75.1 & $\pm$  8.2 \\
    27 &Euterpe      &   108.5 & $\pm$  0.5&  2&   114.1 & $\pm$ 12.2&   111.9 & $\pm$ 10.2&    .... &            \\
    29 &Amphitrite   &   201.0 & $\pm$  3.8&  4&   189.6 & $\pm$ 19.0&   204.9 & $\pm$ 14.6&   212.2 & $\pm$ 22.2 \\
    36 &Atalante     &   108.0 & $\pm$  5.0&  2&   106.0 & $\pm$ 13.8&   118.1 & $\pm$  8.6&   105.6 & $\pm$ 11.4 \\
    38 &Leda         &    97.3 & $\pm$  4.1&  3&    96.4 & $\pm$  9.8&   107.6 & $\pm$  9.8&   115.9 & $\pm$ 11.8 \\
    39 &Laetitia     &   159.0 & $\pm$  1.0&  2&   178.2 & $\pm$ 17.8&   154.0 & $\pm$ 11.0&   149.5 & $\pm$ 17.4 \\
    41 &Daphne       &   185.3 & $\pm$  5.9&  6&   205.5 & $\pm$ 20.6&   181.2 & $\pm$ 16.2&   174.0 & $\pm$ 21.0 \\
    42 &Isis         &    96.0 & $\pm$  5.0&  2&   111.0 & $\pm$ 11.4&   118.3 & $\pm$  8.6&   100.2 & $\pm$ 14.6 \\
    43 &Ariadne      &    61.7 & $\pm$  4.4&  7&    71.4 & $\pm$  7.4&    54.6 & $\pm$  5.0&    65.9 & $\pm$  9.8 \\
    45 &Eugenia      &   186.5 & $\pm$  3.2& 10&   202.2 & $\pm$ 20.2&   193.5 & $\pm$ 13.8&   214.6 & $\pm$ 21.8 \\
    48 &Doris        &   209.5 & $\pm$  3.5&  2&   215.6 & $\pm$ 22.2&   225.4 & $\pm$ 15.8&   221.8 & $\pm$ 23.4 \\
    51 &Nemausa      &   144.3 & $\pm$  5.5& 20&   138.1 & $\pm$ 13.8&   151.4 & $\pm$ 10.6&   147.9 & $\pm$ 15.0 \\
    52 &Europa       &   314.4 & $\pm$  7.5& 13&   304.7 & $\pm$ 30.6&   355.0 & $\pm$ 25.0&   302.5 & $\pm$ 30.6 \\
    54 &Alexandra    &   140.5 & $\pm$  1.5&  2&   159.7 & $\pm$ 16.2&   144.1 & $\pm$ 10.2&   165.8 & $\pm$ 17.0 \\
    56 &Melete       &   113.4 & $\pm$  4.0& 12&   121.0 & $\pm$ 12.2&   106.3 & $\pm$  7.4&   113.2 & $\pm$ 11.4 \\
    62 &Erato        &    90.5 & $\pm$  0.5&  2&   106.8 & $\pm$ 10.6&    82.0 & $\pm$  5.0&    95.4 & $\pm$  9.8 \\
    71 &Niobe        &    83.5 & $\pm$  0.5&  2&    79.6 & $\pm$  8.2&    86.4 & $\pm$  6.2&    83.4 & $\pm$  8.6 \\
    76 &Freia        &   172.3 & $\pm$ 10.1&  6&   145.5 & $\pm$ 14.6&   174.9 & $\pm$ 12.6&   183.7 & $\pm$ 19.0 \\
    80 &Sappho       &    68.7 & $\pm$  5.7&  3&    69.2 & $\pm$  7.0&    74.0 & $\pm$  6.6&    78.4 & $\pm$  8.2 \\
    85 &Io           &   165.2 & $\pm$  4.8& 10&   167.8 & $\pm$ 17.4&   158.3 & $\pm$ 11.4&   154.8 & $\pm$ 15.8 \\
    87 &Sylvia       &   267.1 & $\pm$ 17.4& 11&   253.1 & $\pm$ 25.4&   261.9 & $\pm$ 18.6&   260.9 & $\pm$ 29.4 \\
    88 &Thisbe       &   218.2 & $\pm$  7.4&  9&   184.3 & $\pm$ 36.6&   216.9 & $\pm$ 15.4&   200.6 & $\pm$ 20.6 \\
    89 &Julia        &   138.0 & $\pm$  1.0&  2&   144.9 & $\pm$ 14.6&   147.0 & $\pm$ 10.6&   151.5 & $\pm$ 15.4 \\
    93 &Minerva      &   162.3 & $\pm$  1.7&  3&   154.1 & $\pm$ 15.4&   149.1 & $\pm$ 10.6&   141.6 & $\pm$ 14.6 \\
    94 &Aurora       &   193.5 & $\pm$  6.2& 13&   173.9 & $\pm$ 17.8&   183.3 & $\pm$ 16.6&   204.9 & $\pm$ 21.0 \\
    95 &Arethusa     &   147.0 & $\pm$  1.4&  3&   148.3 & $\pm$ 15.4&   147.3 & $\pm$ 21.0&   136.0 & $\pm$ 17.0 \\
    99 &Dike         &    69.0 & $\pm$  3.0&  2&    69.8 & $\pm$  7.0&    73.2 & $\pm$  5.4&    69.0 & $\pm$ 10.2 \\
   107 &Camilla      &   225.4 & $\pm$ 19.8&  8&   210.4 & $\pm$ 22.6&   200.8 & $\pm$ 18.2&   222.6 & $\pm$ 28.2 \\
   109 &Felicitas    &    84.5 & $\pm$  2.7&  4&    87.3 & $\pm$  8.6&    84.2 & $\pm$ 11.8&    89.4 & $\pm$  9.4 \\
   116 &Sirona       &    76.0 & $\pm$  3.0&  2&      ....  &        &    75.1 & $\pm$  6.6&    71.7 & $\pm$  9.4 \\
   120 &Lachesis     &   158.0 & $\pm$  4.0&  2&   155.3 & $\pm$ 15.8&   170.7 & $\pm$ 11.4&   174.1 & $\pm$ 17.8 \\
   121 &Hermione     &   191.5 & $\pm$  0.5&  2&   167.4 & $\pm$ 19.0&   185.1 & $\pm$ 13.4&   209.0 & $\pm$ 21.4 \\
   129 &Antigone     &   126.9 & $\pm$  4.1&  7&   128.7 & $\pm$ 13.0&   130.2 & $\pm$ 11.8&       .... &         \\
   130 &Elektra      &   192.4 & $\pm$  2.3&  7&   174.4 & $\pm$ 17.8&   182.5 & $\pm$ 13.0&   182.3 & $\pm$ 21.8 \\
   135 &Hertha       &    79.3 & $\pm$  2.5&  3&    71.0 & $\pm$  7.4&    74.5 & $\pm$  5.4&    79.2 & $\pm$  8.2 \\
   144 &Vibilia      &   143.7 & $\pm$  3.3& 11&   155.3 & $\pm$ 27.4&   146.7 & $\pm$ 10.6&   142.4 & $\pm$ 14.6 \\
   146 &Lucina       &   124.9 & $\pm$  4.8&  7&   153.4 & $\pm$ 15.4&   131.6 & $\pm$ 11.8&   132.2 & $\pm$ 13.4 \\
   154 &Bertha       &   167.3 & $\pm$  1.7&  3&   186.1 & $\pm$ 18.6&   187.9 & $\pm$ 17.0&   184.9 & $\pm$ 19.0 \\
   159 &Aemilia      &   132.0 & $\pm$  2.0&  2&   125.2 & $\pm$ 12.6&   138.1 & $\pm$ 12.6&   125.0 & $\pm$ 12.6 \\
   165 &Loreley      &   160.0 & $\pm$ 16.3&  4&   180.0 & $\pm$ 18.2&   185.8 & $\pm$ 13.4&   154.8 & $\pm$ 16.2 \\
   166 &Rhodope      &    50.0 & $\pm$  3.6&  5&    53.1 & $\pm$  5.4&    52.4 & $\pm$  3.0&    .... &            \\
   187 &Lamberta     &   141.5 & $\pm$  0.5&  2&   143.0 & $\pm$ 14.2&   133.1 & $\pm$  9.4&   130.4 & $\pm$ 13.4 \\
   192 &Nausikaa     &    95.7 & $\pm$  1.7&  3&    97.8 & $\pm$  9.8&    96.5 & $\pm$  7.0&   103.3 & $\pm$ 10.6 \\
   199 &Byblis       &    53.0 & $\pm$  5.0&  2&    76.9 & $\pm$  7.8&    56.5 & $\pm$  8.2&    .... &            \\
   208 &Lacrimosa    &    44.5 & $\pm$  0.5&  2&    40.5 & $\pm$  4.2&    39.4 & $\pm$  3.4&    41.3 & $\pm$  4.6 \\
   216 &Kleopatra    &   117.2 & $\pm$ 13.2& 10&   112.0 & $\pm$ 11.4&   132.0 & $\pm$ 11.8&   135.1 & $\pm$ 13.8 \\
   230 &Athamantis   &   114.0 & $\pm$  4.0&  2&   111.3 & $\pm$ 11.4&   112.9 & $\pm$  8.2&   109.0 & $\pm$ 11.0 \\
   233 &Asterope     &   102.5 & $\pm$  1.5&  2&    99.2 & $\pm$ 10.2&    93.4 & $\pm$  6.6&   102.8 & $\pm$ 13.0 \\
   238 &Hypatia      &   141.7 & $\pm$  9.3&  3&   151.7 & $\pm$ 15.4&   157.6 & $\pm$ 11.0&   148.5 & $\pm$ 15.4 \\
   247 &Eukrate      &   143.0 & $\pm$  0.8&  3&   130.9 & $\pm$ 13.0&   153.7 & $\pm$ 11.0&   134.4 & $\pm$ 13.8 \\
   276 &Adelheid     &   121.3 & $\pm$  5.0&  3&   110.5 & $\pm$ 11.4&   134.2 & $\pm$  9.4&   121.6 & $\pm$ 14.6 \\
   328 &Gudrun       &   103.3 & $\pm$  2.5&  3&   143.1 & $\pm$ 14.2&   126.8 & $\pm$  9.0&   122.9 & $\pm$ 13.4 \\
   329 &Svea         &    73.3 & $\pm$  2.5&  3&    80.7 & $\pm$  8.2&    75.2 & $\pm$  5.4&    77.8 & $\pm$  7.8 \\
   334 &Chicago      &   179.3 & $\pm$  2.9&  4&   198.8 & $\pm$ 20.6&   180.1 & $\pm$ 12.6&   158.6 & $\pm$ 18.2 \\
   336 &Lacadiera    &    71.5 & $\pm$  5.5&  2&    66.3 & $\pm$  8.6&    71.0 & $\pm$  5.0&    69.3 & $\pm$  7.4 \\
   345 &Tercidina    &   101.5 & $\pm$  4.5&  2&    96.9 & $\pm$ 12.6&   101.1 & $\pm$  7.4&    94.1 & $\pm$ 10.6 \\
   347 &Pariana      &    52.3 & $\pm$  4.5&  3&    48.6 & $\pm$  5.0&    52.3 & $\pm$  3.8&    51.4 & $\pm$  7.0 \\
   349 &Dembowska    &   149.0 & $\pm$  1.0&  2&   131.7 & $\pm$ 14.2&   176.2 & $\pm$ 15.8&   139.8 & $\pm$ 14.6 \\
   350 &Ornamenta    &   119.0 & $\pm$  5.0&  2&   127.3 & $\pm$ 13.0&   122.7 & $\pm$  8.6&   118.4 & $\pm$ 12.6 \\
   354 &Eleonora     &   166.0 & $\pm$  5.0&  2&   149.0 & $\pm$ 15.0&   151.0 & $\pm$ 13.4&   155.2 & $\pm$ 17.8 \\
   365 &Corduba      &    94.5 & $\pm$  0.5&  2&    86.8 & $\pm$  8.6&   105.4 & $\pm$  7.4&   105.9 & $\pm$ 11.0 \\
   372 &Palma        &   183.2 & $\pm$ 13.1&  6&   175.8 & $\pm$ 17.8&   190.8 & $\pm$ 17.0&   188.6 & $\pm$ 19.0 \\
   380 &Fiducia      &    73.0 & $\pm$  1.0&  2&    67.7 & $\pm$  7.4&    75.8 & $\pm$  7.0&    73.2 & $\pm$  7.8 \\
   386 &Siegena      &   175.5 & $\pm$  2.1&  4&   220.4 & $\pm$ 37.8&   192.7 & $\pm$ 27.4&   165.0 & $\pm$ 16.6 \\
   404 &Arsinoe      &    95.7 & $\pm$  3.3&  3&   100.9 & $\pm$ 10.2&    92.1 & $\pm$  6.6&    97.7 & $\pm$  9.8 \\
   409 &Aspasia      &   159.9 & $\pm$ 15.2&  7&   181.8 & $\pm$ 18.2&   201.7 & $\pm$ 28.6&   161.6 & $\pm$ 17.4 \\
   419 &Aurelia      &   123.0 & $\pm$  2.0&  2&   148.4 & $\pm$ 15.0&   126.1 & $\pm$ 11.4&   129.0 & $\pm$ 13.4 \\
   423 &Diotima      &   205.7 & $\pm$  8.4&  3&   176.1 & $\pm$ 18.2&   224.5 & $\pm$ 20.6&   208.8 & $\pm$ 21.4 \\
   426 &Hippo        &   113.3 & $\pm$  5.8&  4&   128.1 & $\pm$ 13.0&   127.6 & $\pm$ 13.0&   127.1 & $\pm$ 13.4 \\
   433 &Eros         &    17.0 & $\pm$  5.0&  3&    .... &           &    16.5 & $\pm$  1.8&    .... &            \\
   458 &Hercynia     &    37.0 & $\pm$  1.0&  2&    36.6 & $\pm$  3.8&    42.8 & $\pm$  5.0&    38.8 & $\pm$  4.2 \\
   468 &Lina         &    65.7 & $\pm$  1.9&  3&    61.3 & $\pm$  6.2&    58.3 & $\pm$  5.4&    69.3 & $\pm$  7.4 \\
   471 &Papagena     &   126.4 & $\pm$  6.4&  5&   148.1 & $\pm$ 15.4&   118.2 & $\pm$  8.6&   134.2 & $\pm$ 14.6 \\
   489 &Comacina     &   128.0 & $\pm$  4.3&  3&   117.9 & $\pm$ 25.8&   137.8 & $\pm$  9.8&   139.4 & $\pm$ 14.2 \\
   490 &Veritas      &   121.5 & $\pm$  7.5&  2&   118.6 & $\pm$ 12.2&   108.7 & $\pm$  7.8&   115.6 & $\pm$ 13.0 \\
   578 &Happelia     &    69.5 & $\pm$  1.5&  2&    66.7 & $\pm$  7.4&    68.5 & $\pm$  6.2&    69.3 & $\pm$  7.4 \\
   580 &Selene       &    49.5 & $\pm$  0.5&  2&    47.7 & $\pm$  5.0&    48.5 & $\pm$  4.2&    45.8 & $\pm$  5.8 \\
   675 &Ludmilla     &    74.4 & $\pm$  3.9&  7&    .... &           &    65.3 & $\pm$  6.6&    .... &            \\
   694 &Ekard        &   101.3 & $\pm$  2.5&  3&   121.8 & $\pm$ 12.2&    93.3 & $\pm$  6.6&    90.8 & $\pm$  9.8 \\
   695 &Bella        &    49.5 & $\pm$  1.5&  2&    40.5 & $\pm$  4.2&    39.2 & $\pm$  3.0&    48.2 & $\pm$  5.0 \\
   747 &Winchester   &   176.8 & $\pm$ 11.8&  5&   174.9 & $\pm$ 37.4&   165.8 & $\pm$ 13.8&   171.7 & $\pm$ 17.4 \\
   757 &Portlandia   &    36.0 & $\pm$  5.0&  2&    32.9 & $\pm$  3.4&    36.0 & $\pm$  3.4&    32.1 & $\pm$  3.4 \\
   762 &Pulcova      &   141.0 & $\pm$  5.0&  3&   140.1 & $\pm$ 14.2&   132.2 & $\pm$  9.4&   137.1 & $\pm$ 14.2 \\
   791 &Ani          &    80.0 & $\pm$  5.6&  4&   116.6 & $\pm$ 11.8&    92.9 & $\pm$  6.2&   103.5 & $\pm$ 10.6 \\
   834 &Burnhamia    &    65.0 & $\pm$  2.8&  3&    61.3 & $\pm$  6.2&    63.3 & $\pm$ 12.6&    66.7 & $\pm$  7.0 \\
   925 &Alphonsina   &    58.0 & $\pm$  0.8&  3&    57.5 & $\pm$  5.8&    65.2 & $\pm$  5.4&    54.3 & $\pm$  6.6 \\
  1437 &Diomedes     &   129.0 & $\pm$  9.0&  2&   117.8 & $\pm$ 11.8&   165.5 & $\pm$ 13.4&   164.3 & $\pm$ 17.0 \\
  3200 &Phaethon     &     5.3 & $\pm$  0.4&  4&    .... &           &     4.8 & $\pm$  1.0&     5.1 & $\pm$  0.6 
    \label{Diameters}
\end{longtable}
\end{center}
\endgroup
\twocolumn
\nopagebreak

\section*{Data Availability}

The data-set of asteroidal occultation observations can be browsed or downloaded from NASA's Planetary Data System, Small Bodies Node\textsuperscript{\ref{PDS}}.

The working file of observations (which is regularly updated) together with its format specification, can be downloaded from \textcolor{blue}{http://www.lunar-occultations.com/occult4/asteroid\_observations.zip}

Asteroid light curve variations are available in the Asteroid Light Curve Database\textsuperscript{\ref{LCDB}} \citep{lcdb}, in the file LC\_SUM\_PUB.TXT held in LCLIST\_PUB\_CURRENT.zip.

Asteroid light curves from occultations can be accessed at VizieR, catalogue B/occ\textsuperscript{\ref{VizieR}}.

\bibliographystyle{mnras}
\bibliography{asteroids} 

\bsp	
\label{lastpage}
\end{document}